\documentclass[12pt]{article} 
\usepackage[utf8x]{inputenc} 
\usepackage[T1]{fontenc} 
\usepackage{amsmath,amssymb,bbm,bm,commath,mathtools,wasysym,xfrac} 
\usepackage{xcolor} 
\usepackage{graphicx} 
\usepackage{multirow}
\usepackage{cite} 

\usepackage{sectsty} 
\usepackage{geometry} 
\usepackage{fancyhdr} 
\usepackage{setspace} 
\usepackage[nottoc,notlof,notlot]{tocbibind} 
\usepackage[titles]{tocloft} 
\usepackage[unicode]{hyperref} 
\hypersetup{bookmarksnumbered=true, bookmarksopen=true, bookmarksopenlevel=2, breaklinks=true, citecolor=blue, colorlinks=true, linkcolor=red, linktoc=page, pdfborder={0 0 0}, pdfstartview=FitH, plainpages=false, unicode=true, urlcolor=cyan}

\DeclareUnicodeCharacter{"0393}{\Gamma}
\DeclareUnicodeCharacter{"0394}{\Delta}
\DeclareUnicodeCharacter{"0398}{\Theta}
\DeclareUnicodeCharacter{"039B}{\Lambda}
\DeclareUnicodeCharacter{"039E}{\Xi}
\DeclareUnicodeCharacter{"03A0}{\Pi}
\DeclareUnicodeCharacter{"03A3}{\Sigma}
\DeclareUnicodeCharacter{"03A5}{\Upsilon}
\DeclareUnicodeCharacter{"03A6}{\Phi}
\DeclareUnicodeCharacter{"03A8}{\Psi}
\DeclareUnicodeCharacter{"03A9}{\Omega}
\DeclareUnicodeCharacter{"03B1}{\alpha}
\DeclareUnicodeCharacter{"03B2}{\beta}
\DeclareUnicodeCharacter{"03B3}{\gamma}
\DeclareUnicodeCharacter{"03B4}{\delta}
\DeclareUnicodeCharacter{"03B5}{\epsilon}
\DeclareUnicodeCharacter{"03B6}{\zeta}
\DeclareUnicodeCharacter{"03B7}{\eta}
\DeclareUnicodeCharacter{"03B8}{\theta}
\DeclareUnicodeCharacter{"03D1}{\vartheta}
\DeclareUnicodeCharacter{"03B9}{\iota}
\DeclareUnicodeCharacter{"03BA}{\kappa}
\DeclareUnicodeCharacter{"03BB}{\lambda}
\DeclareUnicodeCharacter{"03BC}{\mu}
\DeclareUnicodeCharacter{"03BD}{\nu}
\DeclareUnicodeCharacter{"03BE}{\xi}
\DeclareUnicodeCharacter{"03C0}{\pi}
\DeclareUnicodeCharacter{"03C1}{\rho}
\DeclareUnicodeCharacter{"03C3}{\sigma} 
\DeclareUnicodeCharacter{"03C4}{\tau}
\DeclareUnicodeCharacter{"03C5}{\upsilon}
\DeclareUnicodeCharacter{"03C6}{\phi}
\DeclareUnicodeCharacter{"03D5}{\varphi}
\DeclareUnicodeCharacter{"03C7}{\chi}
\DeclareUnicodeCharacter{"03C8}{\psi}
\DeclareUnicodeCharacter{"03C9}{\omega}
\DeclareUnicodeCharacter{"21D0}{\Leftarrow}
\DeclareUnicodeCharacter{"0212B}{\AA}
\DeclareUnicodeCharacter{"00B7}{\cdot}
\DeclareUnicodeCharacter{"00B0}{^{\circ}}
\DeclareUnicodeCharacter{"266A}{\eighthnote}
\DeclareUnicodeCharacter{"266B}{\twonotes}
\PrerenderUnicode{¹}\PrerenderUnicode{³}
\PrerenderUnicode{⁴}\PrerenderUnicode{⁵}
\PrerenderUnicode{×}\PrerenderUnicode{̃}

\definecolor{title}{rgb}{0.2,0.5,0.9}
\definecolor{abst}{rgb}{0.366,0.366,0.366}
\definecolor{sect}{rgb}{0.2,0.4,0.7}
\definecolor{ssect}{rgb}{0.4,0.5,1.0}
\definecolor{sssect}{rgb}{0.4,0.6,0.9}
\definecolor{appsect}{rgb}{0.9,0.2,0.5}
\definecolor{ref}{rgb}{0.0,0.0,1.0}
\definecolor{orcidlogocol}{HTML}{A6CE39}
\newcommand{\Title}[1] {\title{\color{title}\Huge #1}}
\newcommand{\TPheader}[3] {\date{}\maketitle\thispagestyle{fancy}\pagenumbering{alph}\lhead{#1}\chead{#2}\rhead{#3}\cfoot{}}

\newcommand{\Abstract}[1] {\begin{abstract}\normalsize #1 \end{abstract}}
\newcommand{\makepage}[1] {\newpage\pagenumbering{#1}}

\sectionfont{\color{sect}}
\subsectionfont{\color{ssect}}
\subsubsectionfont{\color{sssect}}
\renewcommand{\appendix}{\setcounter{section}{0}\sectionfont{\color{appsect}}\renewcommand{\thesection}{\Alph{section}}\renewcommand*{\theHsection}{app.\the\value{section}}} 
\newcommand\references[1]{\sectionfont{\color{ref}}\bibliographystyle{hephys}\bibliography{#1}}

\newcommand\eqs[1] {\begin{align}#1\end{align}}

\newcommand\eqsc[1] {\begin{gather}#1\end{gather}}

\newcommand\eqsa[1] {\equ{\begin{aligned}#1\end{aligned}}}
\newcommand\eqsg[1] {\equ{\begin{gathered}#1\end{gathered}}}
\newcommand\equ[1] {\begin{equation}#1\end{equation}}

\newcommand\tabl[2] {\begin{table}[#1]\centering #2\end{table}}

\newcommand\half {\tfrac{1}{2}}

\renewcommand\( {\left(}
\renewcommand\) {\right)}
\renewcommand\Im {\text{Im}}
\renewcommand\Re {\text{Re}}

\newcommand\wt {\widetilde}

\DeclareMathOperator{\Tr}{Tr}

\DeclareMathOperator{\sgn}{sgn}

\newcommand\bC {{\mathbb C}}

\newcommand\bR {{\mathbb R}}

\def\C {{\mathcal C}}

\newcommand\F {{\mathcal F}}
\renewcommand\H {{\mathcal H}}

\newcommand\M {{\mathcal M}}
\newcommand\N {{\mathcal N}} 
\renewcommand\O {{\mathcal O}}
\newcommand\R {{\mathcal R}}
\renewcommand\S {{\mathcal S}}

\newcommand\W {{\mathcal W}}
\newcommand\Z {{\mathcal Z}}
\newcommand\fg {{\mathfrak g}}
\newcommand\fm {{\mathfrak m}}
\newcommand\fn {\mathfrak{n}}
\newcommand\fq {\mathfrak{q}}

\newcommand{\tu}{{\tilde{u}}}
\newcommand{\tnu}{{\tilde{\nu}}}

\newcommand\nn {\nonumber\\}

\numberwithin{equation}{section} 
\begin{document}
\Title{Notes on 5d Partition Functions -- I}

\author{\href{mailto:dharmesh.jain@outlook.com}{Dharmesh Jain}\,\href{https://orcid.org/0000-0002-9310-7012}{\includegraphics[scale=0.0775]{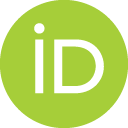}}\\
\emph{\normalsize Department of Theoretical Sciences, S. N. Bose National Centre for Basic Sciences,}\\
\emph{\normalsize Block--JD, Sector--III, Salt Lake City, Kolkata 700106, India}\bigskip\\
}

\TPheader{}{\today}{} 

\Abstract{We continue the study of partition functions of 5d supersymmetric theories on manifolds taking the form of a twisted product $\M_3×Σ_{\fg}$ with $Σ_{\fg}$ denoting a Riemann surface of genus $\fg$. The 5d theory compactified on $Σ_{\fg}$ leads to a novel class of 3d theories in IR, whose existence at large $N$ is expected from holography. Focussing on $\M_3$ being $S^2×S^1$ without or with a topological twist on the 2-sphere leads to the superconformal index or topologically twisted index, respectively, for such a class of 3d theories. We discuss the large $N$ limit of these partition functions and find new relations between them and other well-known 5d partition functions, with interesting consequences for the 3d indices.
}

\tableofcontents
\makepage{arabic} 

\section{Introduction and Summary}
A novel class of 3d $\N=2$ SCFTs arises as a fixed point in the IR when 5d $\N=1$ gauge theories are compactified on a Riemann surface $Σ_{\fg}$ of genus $\fg$ with a partial topological twist. Such theories on squashed 3-sphere ($S^3_b$) were studied extensively in \cite{Crichigno:2018adf} and the exact 5d partition function on $S^3_b×Σ_{\fg}$ was computed. The large $N$ results for the corresponding free energy were shown to match the holographic predictions\cite{Bobev:2017uzs,Bah:2018lyv}, providing evidence for the existence of strongly coupled 3d theories with gravity duals (given that the original 5d theory also has a dual). The 5d theories on $\M_3×Σ_{\fg}$ have also opened up new exploratory avenues in holography, for example, with the prediction of 5d Cardy-like formulae valid in extended regime of fugacities \cite{Bobev:2019zmz}, that mimic the behaviour of 3d Cardy-like formulae of \cite{Choi:2019zpz,Choi:2019dfu}.

In this note, we study the partition functions on 5d manifolds given by $(S^2_{ε_1}×S^1)×Σ_{\fg_2}$ and $(\wt{S^2_{ε_1}}×S^1)×Σ_{\fg_2}$, which lead to the 3d superconformal index and topologically twisted index (SCI and TTI), respectively, when we consider compactification on the Riemann surface $Σ_{\fg_2}$ with vanishing size. We have also introduced refinement $ε_1$ for the 2-sphere with respect to the angular momentum and the $\wt{S^2}$ denotes a partial topological twist on $S^2$.\footnote{There is always such a twist on the Riemann surface so we do not use a tilde on $Σ_{\fg_2}$. Also, we mostly set the radius of $S^1$ to 1.} The general structure of both these partition functions schematically looks like
\equ{Z(\tnu;\fn_1,\fn_2)=\frac{1}{|W_G|}∑_{\fm_1,\fm_2}∮d\tu^i\Z^{cl}(\tu;\fm_{1,2})\Z^{1-loop}(\tu,\tnu;\fm_{1,2},\fn_{1,2})\Z^{inst}(\tu,\tnu;\fm_{1,2},\fn_{1,2})\,,
}
where $\tu^i$ are the Cartan-valued gauge variables, $\tnu$ are the flavour chemical potentials (or complexified supersymmetric mass parameters), $\fm_1(\fn_1)$ are gauge (flavour) fluxes on $S^2$ and $\fm_2(\fn_2)$ are gauge (flavour) fluxes on $Σ_{\fg}$. The gauge fluxes lie in the coweight lattice $Λ_G$ of gauge group $G$ and $|W_G|$ denotes the order of the Weyl group of $G$. Let us now discuss the two partition functions separately.

\paragraph{$\bm{(S^2_{ε_1}×S^1)×Σ_{\fg_2}}$.} This is one of the least studied partition functions in 5d. We will construct it by gluing four 5d Nekrasov partition functions, following the approach used in \cite{Crichigno:2018adf,Hosseini:2018uzp}. These gluing conditions will be such that the 3d ``identity gluing'' of holomorphic blocks to get the 3d SCI \cite{Beem:2012mb,Closset:2018ghr} will be baked in the 5d rules that are, otherwise, similar to those discussed in \cite{Bawane:2014uka}, that has partial topological twist along both $S^2_{ε_1}$ and $Σ_{\fg_2}$. With the partition function in hand, we will study its large $N$ limit for a class of 5d quiver theories (assumed to have massive type IIA string duals) for which the instanton contributions are exponentially suppressed. To do that, we will follow a strategy similar to the evaluation of the $S^3_b×Σ_{\fg}$ partition function at large $N$:
\begin{enumerate}
\item Sum over fluxes $\fm_2$ on $Σ_{\fg}$ to generate simple poles;

\item Extremize the \emph{twisted superpotential} characterizing these poles to get the saddle configuration $\{u^*,\fm_1^*\}$;

\item Use this configuration to evaluate the residues and get the partition function as a function of flavour chemical potentials and fluxes.
\end{enumerate}

We will find that the flux $\fm_1$ on $S^2$ does not contribute at large $N$, or to put it another way, $\fm_1=0$ is the saddle point. In the absence of flavour fluxes (this case corresponds to the so-called universal twist \cite{Benini:2015bwz,Azzurli:2017kxo,Bobev:2017uzs}\footnote{The universal twist corresponds to the partial topological twist performed using the exact UV superconformal R-symmetry and no mixing with the flavour symmetries. Since this twist requires only R-symmetry, it leads to \emph{universal} relations.}), we prove the large $N$ relation predicted in \cite{Bobev:2019zmz}:
\equ{\log Z_{(S^2_{ε_1}×S^1)×Σ_{\fg_2}} =\frac{i}{π}\frac{φ^2}{ω}F_{S^3×Σ_{\fg_2}}\,,
}
where $φ=ω+2πi$ with $ω=2πiε_1$ and $F=-\log Z$ denotes the free energy. Turning the flavour fugacities and fluxes on, we get an interesting formula for $\log Z$ written in terms of the twisted superpotential $\W$ as follows:
\equ{\log Z_{(S^2_{ε_1}×S^1)×Σ_{\fg_2}} = -2πi(\fg_2-1)\bigg[4\W +∑_I(\fn_{2I} -2\tnu_I)\frac{∂\W}{∂\tnu_I} -2ε_1\frac{∂\W}{∂ε_1}\bigg]\,,
\label{5dSCIintro}}
where $\W$ can be written in terms of $F_{S^5}$ as tabulated in Table \ref{tab:relPFs} and derived later in Section \ref{sec:5dUofSCI}.

Jumping ahead, we would like to mention that in the strict Cardy limit ($ε_1≪1$), this formula leads to a factorized form as expected from the 3d Cardy results of \cite{Choi:2019dfu}. But this does not hold in general for finite $ε_1$. This non-factorization behaviour at large $N$ for the novel class of 3d theories obtained as nontrivial IR fixed point of 5d theories compactified on $Σ_{\fg_2}$ with gravity duals ($N^{\frac{5}{2}}$ scaling) differs from that of usual 3d theories with M-theory duals ($N^{\frac{3}{2}}$ scaling) where the Cardy formula seems to be valid for finite $ε_1$ too.

\paragraph{$\bm{(\wt{S^2_{ε_1}}×S^1)×Σ_{\fg_2}}$.} The unrefined version ($ε_1=0$) of this partition function has been studied before in \cite{Crichigno:2018adf,Hosseini:2018uzp}. In both these papers, a conjecture was made regarding the prepotential of the 5d theory determining the eigenvalue density of the large $N$ matrix model, and the final result supported that conjecture. We revisit this calculation for the refined case here and find that the extremization of the prepotential is an avoidable step. We will follow the same steps outlined above that will directly produce the expected result. Though, in this case, we will have nonvanishing contributions from flux $\fm_1$ as expected from \cite{Hosseini:2018uzp}. The evaluation of the prepotential seems to work in the unrefined case because the Nekrasov-Shatashvili limit of the twisted superpotential is proportional to the prepotential in the $ε_1→0$ limit: $\W^{(5d)}_{NS}\sim\frac{1}{ε_1}\F$ (see \cite{Crichigno:2018adf} for more details).

We will find that the general result for the partition function with refinement in terms of the twisted superpotential looks formally the same as \eqref{5dSCIintro}:
\equ{\log Z_{(\wt{S^2_{ε_1}}×S^1)×Σ_{\fg_2}} = -2πi(\fg_2-1)\bigg[4\W +∑_I(\fn_{2I} -2\tnu_I)\frac{∂\W}{∂\tnu_I} -2ε_1\frac{∂\W}{∂ε_1}\bigg]\,,
\label{5dTTIintro}}
but now the $\W$ is directly related to $F_{S^3×Σ_{\fg_2}}$ as shown in Table \ref{tab:relPFs} (or a particular derivative of $F_{S_5}$ as discussed later in Section \ref{sec:5dUofTTI}). For vanishing $ε_1$, this relation matches the relation between 3d TTI and $F_{S^3}$, as one might expect from \cite{Hosseini:2016tor,Jain:2019lqb}. We are not aware of the large $N$ analysis of 3d TTI with refinement in the literature, but we believe that \eqref{5dTTIintro} should hold for pure 3d quiver theories too, with $(\fg_2-1)\W$ being proportional to $F_{S^3}$. The non-factorization behaviour for finite $ε_1$ mentioned above continues to hold here too.

\paragraph{Summary.} We list in Table \ref{tab:relPFs} the large $N$ relations between various 5d partition functions (and twisted superpotentials where necessary) and mass-deformed free energy on round $S^5$, $F_{S^5}(m)$. The first three relations in this table have been derived at various places with various sophistication in the literature, see \cite{Crichigno:2018adf,Crichigno:2020ouj} and references therein for relevant details. The last two relations are new and are derived later in this note.
\tabl{!h}{\begin{tabular}{c|c|c}
No. & Manifold & Relation \\
\hline
1. & $S^5_{(ω_1,ω_2,ω_3)}$ & $F(m)=\frac{(ω_{tot})^3}{27ω_1ω_2ω_3}F_{S^5}\big(\frac{3m}{ω_{tot}}\big)$;\;$\scriptstyle ω_{tot}=ω_1+ω_2+ω_3$\vphantom{\bigg)} \\
\hline
2. & $S^4_{ε_1,ε_2}×S^1_r$ & $\log Z(ν)=-2F_{S^5_{(ε_1,ε_2,r^{-1})}}\(ν\)$ \vphantom{\bigg)} \\
\hline
3. & $S^3_b×Σ_{\fg}$ & $\begin{matrix}\vphantom{\bigg(}\W(\tnu) =\frac{4}{27π}F_{S^5_{(b,b^{-1},Q)}}\big(\frac{3Q\tnu}{2}\big)$;\;$\scriptstyle Q=\frac{1}{2}(b+b^{-1}) \\[2mm] \log Z(\tnu;\fn)= 2π(\fg-1)\left[3\W +∑_I(\fn_I -\tnu_I)\frac{∂\W}{∂\tnu_I}\right]\vphantom{\bigg)}\end{matrix}$ \\
\hline\hline
4. & $(S^2_{ε_1}×S^1)×Σ_{\fg_2}$ & $\begin{matrix}\vphantom{\bigg(}(2πi)\W(\tnu;\fn_1) =-\frac{8(1+ε_1)}{27}F_{S^5_{(\sqrt{ε_1},(\sqrt{ε_1})^{-1},Q)}}\big(\frac{3Q\tnu^{±}}{2}\big)$;\;$\scriptstyle Q=\frac{1+ε_1}{2\sqrt{ε_1}},\;\tnu^{±}=\frac{2\tnu ±ε_1\fn_1}{1+ε_1} \\ \log Z(\tnu;\fn_1,\fn_2)= -2πi(\fg_2-1)\left[4\W +∑_I(\fn_{2I} -2\tnu_I)\frac{∂\W}{∂\tnu_I} -2ε_1\frac{∂\W}{∂ε_1}\right]\vphantom{\bigg)}\end{matrix}$ \\
\hline
5. & $(\wt{S^2_{ε_1}}×S^1)×Σ_{\fg_2}$ & $\begin{matrix}\vphantom{\bigg(}(2πi)(\fg_2-1)\W(\tnu;\fn_1) =\frac{1}{4}F_{S^3_{b=1}×Σ_{\fg_2}}\Big(2\tnu±ε_1\sqrt{\fn_1^2 -1};\fn_1\Big) \\ \log Z(\tnu;\fn_1,\fn_2)= -2πi(\fg_2-1)\left[4\W +∑_I(\fn_{2I} -2\tnu_I)\frac{∂\W}{∂\tnu_I} -2ε_1\frac{∂\W}{∂ε_1}\right]\vphantom{\bigg)} \end{matrix}$ \\
\hline
\end{tabular}
\caption{Summary of relations between various 5d partition functions. For other versions of the relations in 4 and 5, we refer the reader to Sections \ref{sec:5dUofSCI} and \ref{sec:5dUofTTI}, respectively. Note that indices on flavour fugacities and fluxes are suppressed unless required by the presence of summation.}
\label{tab:relPFs}
}

Just for completeness, we recall that for theories with massive type IIA duals\cite{Seiberg:1996bd,Intriligator:1997pq,Bergman:2012kr}, the mass-deformed free energy on round $S^5$ reads\cite{Jafferis:2012iv,Chang:2017mxc,Hosseini:2019and}
\equ{F_{S^5}(m)=\(\frac{∑_Ic_ρ^I\(1-\frac{4}{9}m_I^2\)}{∑_Ic_ρ^I}\)^{\frac{3}{2}}F_{S^5}\,,
\label{FS5mexp}}
where the index $I$ labels various hypermultiplets in the quiver under consideration and the numerical coefficients $c_ρ$ depend on those hypermultiplet's representations, for example, $c_{adj}=c_{antisym}=1$, $c_{bifun}=2$, $c_{fun}=0$ (see \cite{Crichigno:2018adf} for details). The theories with type IIB duals\cite{DHoker:2016ysh,Uhlemann:2019ypp} do not yet have a similar explicit formula with non-zero masses. Though, we believe that the formulae given in the Table \ref{tab:relPFs} should hold for such theories too. Of course, in the case of vanishing masses and chemical potentials, it can be checked that the above relations in 4 and 5 also hold for theories with type IIB duals, following arguments similar to those outlined in \cite{Crichigno:2020ouj}, which cover the relations in 1-3.

\paragraph{Outline.} This note is organized as follows. In Section \ref{sec:Review} we review the basic building blocks of the 5d partition functions, evaluation of the partition functions, including the explicit extremization of the resulting matrix models. In Section \ref{sec:5dUofSCI} we construct the 5d partition function of $(S^2_{ε _1}×S^1)×Σ_{\fg_2}$ and study the large $N$ limit of the partition function for generic quiver theories. We repeat a similar analysis for $(\wt{S^2_{ε_1}}×S^1)×Σ_{\fg_2}$ in Section \ref{sec:5dUofTTI}. We end with a discussion of future outlook in Section \ref{sec:Disc} and a functional Appendix \ref{app:Id}.

\section{Review}\label{sec:Review}
We use the notations and conventions of \cite{Crichigno:2018adf,Crichigno:2020ouj} (see also \cite{Hosseini:2018uzp}) most of the time. For orientation purposes, we review a few necessary ingredients below.

\subsection{5d Nekrasov Partition Function}\label{sec:NPFs}
The path integral on various 5d manifolds localizes around certain fixed points where the local geometry looks like a copy of $\Omega$-deformed background $\bC^{2}_{ε_{1},ε_{2}}\times S^{1}_{r}$. The instanton partition function \cite{Nekrasov:2002qd} on this background then serves as a basic building block for various 5d partition functions. The 5d partition function is obtained by appropriately gluing copies of the instanton partition function and integrating/summing over gauge configurations:
\equ{Z_{M_{5}} =\frac{1}{|W_G|}\sum_{\fm∈Λ_G}\oint_{\C} \frac{dx}{2\pi i x}\prod_{\ell} \Z^{pert}_{\bC^2\times S^1}(x^{(\ell)},y^{(\ell)};\fq_1^{(\ell)};\fq_2^{(\ell)})\Z^{inst}_{\bC^2\times S^1}(x^{(\ell)},y^{(\ell)};\fq_1^{(\ell)};\fq_2^{(\ell)})\,,
\label{gencopies}}
where $x=e^{2\pi i r \tu}$, $y=e^{2\pi i r \tnu}$, and $\fq_i=e^{2\pi i r ε_i}$. The precise gluing conditions on the parameters, and the summation/integration over gauge variables and the contour $\C$ depend on the specific choice of the 5-manifold. For the manifolds we consider in this note, we will also take the limit $ε_2→0$.

The perturbative part \cite{Bershtein:2018srt} consists of a classical and 1-loop contributions,
\eqs{\Z^{pert}_{\bC^2\times S^1} &=\Z_{\bC^2\times S^1}^{cl}\Z_{\bC^2\times S^1}^{1-loop}\,, \nn
\Z^{cl}_{\bC^2_{\vec{ε}}\times S^1_r}(x;\fq_1,\fq_2) &=\text{exp}\(\frac{8π^3r}{g^2ε_1ε_2}\Tr _F(\tu^2) +\frac{i π k r}{3ε_1ε_2}\Tr_F(\tu^3)\),
}
and the 1-loop contributions from vector multiplet and hypermultiplet are given by
\eqsa{\Z^{1-loop, vec}_{\bC^2_{\vec{ε}}\times S^1_r}(x;\fq_1,\fq_2) &=\prod_{\alpha\in Ad(G)'} \left[(x^{\alpha}\, \fq_{1}\fq_{2};\fq_{1},\fq_{2})(x^{-\alpha};\fq_{1},\fq_{2})\right]^{\frac{1}{2}}\,, \\
\Z^{1-loop, hyp}_{\bC^2_{\vec{ε}}\times S^1_r}(x,y;\fq_1,\fq_2) &=\prod_{\rho\in \R}\left[\big(-x^{\rho}y \sqrt{\fq_{1}\fq_{2}};\fq_{1},\fq_{2}\big)\big(-x^{-\rho}y^{-1} \sqrt{\fq_{1}\fq_{2}};\fq_{1},\fq_{2}\big)\right]^{-\frac{1}{2}}\,,
\label{5dblock}}
where $α$ and $ρ$ denote roots and weights, respectively, such that $x^{α}=e^{2\pi i r α(\tilde u)}$ and similarly for $x^ρ$. Note the subtle $-$ sign in the hyper $\Z$ that comes due to the shift of KK momenta when uplifting from 4d to 5d. The vector $\Z$ follows from the hyper $\Z$ by inverting and setting $y=-\sqrt{\fq_{1}\fq_{2}}$ as per the definitions of the relevant indices \cite{Kim:2012qf}.

The double-Pochhammer symbol is defined as
\equ{(z;\fq_{1},\fq_{2}) =\prod_{k_{1},k_{2}\geq0}(1-z\fq_{1}^{k_{1}}\fq_{2}^{k_{2}})\,\qquad |\fq_{1}|<1\,,\quad |\fq_{2}|<1
\label{dPdef}}
and satisfies the following ``inversion'' identities
\eqsg{(z;\fq_1,\fq_2)=(z\fq_1^{-1};\fq_1^{-1},\fq_2)^{-1}\,,\quad (z;\fq_1,\fq_2)=(z\fq_1^{-1}\fq_2^{-1};\fq_1^{-1},\fq_2^{-1})\,, \\
(\fq_1z;\fq_1,\fq_2)=(z;\fq_2)^{-1}(z;\fq_1,\fq_2)\,,\quad (\fq_2z;\fq_1,\fq_2)=(z;\fq_1)^{-1}(z;\fq_1,\fq_2)\,,
\label{InvId}}
where $(z;\fq_i)≡(z;\fq_i)_∞$ is the $q$-Pochhammer symbol defined in the Appendix \ref{app:Id}.

The instanton contribution can be schematically written as
\equ{\Z^{inst}_{\bC^2_{\vec{ε}}\times S^1_r}(x,y,z;\fq_1,\fq_2) =∑_{k=0}^{∞}z^k\Z^{(k)}_{\bC^2_{\vec{ε}}\times S^1_r}(x,y;\fq_1,\fq_2)\,,\qquad z=e^{-\frac{16π^3r}{g^2}}\,,
}
where $\Z^{(k)}_{\bC^2_{\vec{ε}}\times S^1_r}(x,y;\fq_1,\fq_2)$ is the 1-loop determinant in the $k$-instanton background for vector and hypermultiplets\cite{Nekrasov:2002qd,Nekrasov:2003rj,Nekrasov:2004vw}. For the partition functions we are interested in, we need to take $ε_2→0$ limit, which is known as the Nekrasov-Shatashvili (NS) limit of the instanton partition function \cite{Nekrasov:2009rc} and can be explicitly characterized following \cite{Crichigno:2018adf} as:
\equ{\Z^{inst}_{\bC^2_{\vec{ε}}\times S^1_r}(x,y,z;\fq_1,\fq_2) \xrightarrow{ε_2→0} e^{2π i\(\frac{1}{ε_2}\W_{NS}^{(5d),inst}(\tu,\tnu,z;ε_1) -\frac{1}{2}Ω_{NS}^{(5d),inst}(\tu,\tnu,z;ε_1)+\O(ε_2)\)}\,,
}
where $\W$ and $Ω$ refer to twisted superpotential and dilaton (defined in the next subsection). We know from the analysis of \cite{Jafferis:2012iv,Choi:2019miv} that for a large class of 5d theories \cite{Seiberg:1996bd,Bergman:2012kr}, there is a large $N$ regime where instanton contributions are exponentially suppressed (owing mainly to the presence of $z$ and loop corrections to the YM coupling $g$). Since we will mostly focus on large $N$ analysis of the 5d partition functions, we will refrain from writing explicit expressions for the instanton contributions and refer readers to \cite{Kim:2012qf,Kim:2013nva}.

\subsection{Bethe Ansatz Equations and A-twist Formalism}\label{sec:AtFnBAEs}
The final result for the 5d partition functions from localization (including non-perturbative contributions) results in an expression with a schematic form as follows:
\eqsg{Z_{\M_{3} \times \Sigma_\fg}(\tnu;\fn) = \frac{1}{|W_G|} \sum_{\fm \in \Lambda_G} \oint_{\C_{JK}} d\tu\; \Pi_i(\tu,\tnu)^{\fn_i} \Pi_a(\tu,\tnu)^{\fm_a} \H(\tu,\tnu)^\fg e^{-2 \pi i \Omega(\tu,\tnu)}\,, \label{jkformula} \\
\Pi_{a(i)}(\tu,\tnu) = \exp \bigg(2 \pi i \frac{\partial \W_{\M_3 \times \bR^2}}{\partial \tu_a(\tnu_i)} \bigg)\,,\quad \H(\tu,\tnu) = e^{2 \pi i \Omega_{\M_3 \times \bR^2}} \det_{a,b} \frac{\partial^2 \W_{\M_3 \times \bR^2}}{\partial \tu_a \partial \tu_b}\,,
}
where the functions $\W,Ω$ will be identified below, the sum is over gauge fluxes $\fm$ in the coweight lattice $\Lambda_G$, and $\C_{JK}$ is the so-called ``Jeffrey-Kirwan contour'' \cite{Jeffrey:1995,Benini:2013xpa}. We refer to \cite{Closset:2017zgf} for details on the JK contour and the argument that the sum over $\fm_a$ is a convergent geometric series on this contour. After performing this sum, we find
\equ{Z_{\M_{3} \times \Sigma_\fg}(\tnu;\fn) = \frac{1}{|W_G|} \oint_{\C} d\tu\; \Pi_i(\tu,\tnu)^{\fn_i} \prod_a \frac{1}{1-\Pi_a(\tu,\tnu)} \H(\tu,\tnu)^\fg e^{-2 \pi i \Omega(\tu,\tnu)}\,.
}
The poles of this integrand are at the solutions to the following Bethe-Ansatz-type equations
\equ{\S_{\mathit{BE}} = \left\{ \hat{u} \;\; \big| \;\; \Pi_a(\hat{u}) \equiv \exp \bigg( 2 \pi i \frac{\partial \W_{\M_3 \times \bR^2}}{\partial \tu_a}(\hat{u})\bigg) = 1, \;\;\; a=1,...,r_G\right\} /W_G\,,
\label{SBEdefintro}}
appearing with multiplicity $|W_G|$, which cancels the prefactor, and taking their residues we get the final formula
\equ{Z_{\M_3\times \Sigma_\fg}(\tnu;\fn)   = \sum_{\hat{u} \in \S_{\mathit{BE}}} \Pi_i(\hat{u},\tnu)^{\fn_i} \H(\hat{u},\tnu)^{\fg-1}\,.
\label{BEsumintro}}
Thus, to derive the partition function on $\M_{3} \times \Sigma_\fg$ for general $\Sigma_\fg$, it suffices to compute the objects $\W(\tu,\tnu)$ and $\Omega(\tu,\tnu)$.

These objects can be expressed as observables in a certain 2d topological quantum field theory (TQFT), namely the topological A-twist \cite{witten:1988xj} of the effective theory obtained by compactification on $\M_3$ of the 5d parent theory. This is related to the gauge-Bethe correspondence of \cite{Nekrasov:2009uh,Nekrasov:2014xaa}, and can be described as a ``higher dimensional A-twist'' \cite{Closset:2017zgf}. Compactifying a 5d $\N=1$ theory on $\M_{3} \times \bR^2$ leads to an effective 2d $\N=(2,2)$ theory. The result for the partition function on $\M_{3}×Σ_{\fg}$ then takes the form of a sum over supersymmetric ``Bethe vacua'' of this 2d theory as given in \eqref{BEsumintro}, and the objects $Π,\H$ are then referred to as the ``flux operator'' and ``handle-gluing operator'', respectively. They are built in terms of the {\it effective twisted superpotential}, $\W_{\M_3 \times \bR^2}$, and the {\it effective dilaton}, $\Omega_{\M_3 \times\bR^2}$, controlling the low energy effective theory on the 2d Coulomb branch. Finally, the set of supersymmetric Bethe vacua of this theory, $\S_{\mathit{BE}}$, is defined as the solutions to the same Bethe-Ansatz-type equations appearing above in \eqref{SBEdefintro}.

\subsection{Extremization of the Matrix Models}\label{DanVT}
Since we are interested in the large $N$ limit of the 5d partition functions, the most crucial step is to solve \eqref{SBEdefintro}. It is more or less equivalent to extremizing $\W$ with respect to $\tu_a$ and as we will see later, $\W(\tu_a)$ reduces to a nonlocal function $\W(x)$ in the large $N$ limit given by $\tu_a\sim N^{α}x$, with $α>0$ and a continuous variable $x∈[0,x^*]$. In this section, we present a generic solution to this extremization problem for a class of nonlocal `Lagrangians' that arise out of the matrix models for the partition functions we consider in the following sections. The solution is similar in spirit to the ones dealt in \cite{Gulotta:2011si,Jafferis:2012iv,Uhlemann:2019ypp} and can be termed a ``virial theorem'' as done in \cite{Gulotta:2011si}.

Let us start with a general form of the matrix model Lagrangian (could be free energy or twisted superpotential depending on context) dependent on the ``eigenvalue density'' $ρ(x)$ and a parameter $η$ (which will turn out to be related to magnetic fluxes):
\equ{\tilde{F}=-A(η)∫dxdyρ(x)ρ(y)f(x,y) +B(η)∫dx ρ(x)V(x) +μ\(∫dxρ(x) -1\).
\label{Fdef}}
We have arbitrary coefficient functions $A(η)$, $B(η)$ depending on the variable $η$ and $f(x,y)=|x-y|+|x+y|$ such that $∂_xf(x,y)=\sgn(x-y)+1\;⇒\; ∂_x^2f(x,y)=2δ(x-y)$. We actually only need $∂_x^2f(x,y)=2δ(x-y)$ to hold true so the form of $f(x,y)$ can be relaxed a little. Next, we assume $V(x)$ to be a homogeneous function of degree $n$ such that $xV'(x)=nV(x)$, and $V(0)=V'(0)=0$. (We deal with $n=3$ in the following sections.) Also useful is the relation $xV''(x)=(n-1)V'(x)$. The last term contains the Lagrange multiplier $μ$, which enforces the normalization of $ρ(x)$. The equations of motion then follow:
\begingroup
\allowdisplaybreaks
\eqs{\frac{∂\tilde{F}}{∂ρ(x)}=0 \quad &⇒\quad -2A∫dyρ(y)f(x,y) +BV(x) +μ=0 \label{Fext1} \\
\frac{∂\tilde{F}}{∂η}=0 \quad &⇒\quad -A'∫dxdyρ(x)ρ(y)f(x,y) +B'∫dxρ(x)V(x) =0 \label{Fextn} \\
\frac{∂\tilde{F}}{∂μ}=0 \quad &⇒\quad ∫dxρ(x) =1\,. \label{Fext2}
}
\endgroup

Differentiating \eqref{Fext1} with respect to $x$ twice, we get
\begingroup
\allowdisplaybreaks
\eqs{∂_x &:\quad -2A∫dyρ(y)∂_xf(x,y) +BV'(x)=0 \label{Fext3} \\
∂_x^2 &:\quad -4Aρ(x) +BV''(x)=0 \quad ⇒\quad ρ(x)=\frac{BV''(x)}{4A}\,· \label{Fext4}
}
\endgroup
Since $ρ(x)\propto V''(x)$, we get some useful relations:
\equ{xρ(x)=(n-1)\frac{BV'(x)}{4A} \quad \text{ and } \quad (xρ(x))'=(n-1)ρ(x)\,,
}
which will come in handy below. Now, plugging \eqref{Fext4} in \eqref{Fext2}, we ensure normalizability of $ρ(x)$ and get an equation to solve for $x^*$:
\equ{V'(x^*)=\frac{4A}{B}\quad \text{or} \quad BV(x^*)=\frac{4Ax^*}{n}\,·
}
Then, plugging \eqref{Fext4} in \eqref{Fext1}, we can solve for $μ$ after a couple of integration by parts:
\equ{μ=(n-1)BV(x^*)=\frac{4A(n-1)x^*}{n}\quad ⇒\quad \boxed{BV\(\frac{nμ}{4A(n-1)}\)=\frac{μ}{n-1}}\,·
\label{solmu}}

Next, we massage \eqref{Fext3} by multiplying it with $xρ(x)$ and integrating over $x$ to get
\eqsc{2A(n-1)∫dxdyρ(x)ρ(y)f(x,y) -(n-1)\(μ+BV(x^*)\) +nB∫dx ρ(x)V(x) =0 \nn
⇒ (2n-1)B∫dx ρ(x)V(x) =(n-1)BV(x^*) \nn
⇒ B∫dx ρ(x)V(x) =\frac{μ}{2n-1}\,,
\label{Fext5}}
where we used \eqref{Fext1} to replace the nonlocal term in the second line and used \eqref{solmu} to get the third line. Similarly, massaging \eqref{Fext1} we can evaluate the nonlocal integral in terms of $μ$:
\eqsc{-2A∫dxdyρ(x)ρ(y)f(x,y) +\frac{μ}{2n-1} +μ =0 \nn
⇒ A∫dxdyρ(x)ρ(y)f(x,y) =\frac{n\,μ}{2n-1}\,·
\label{FextNL}}
Finally, substituting \eqref{FextNL}, \eqref{Fext5} and \eqref{Fext2} back in \eqref{Fdef}, we get the ``on-shell'' Lagrangian $\bar{F}$:
\equ{\boxed{\bar{F}=-\frac{n-1}{2n-1}μ}\,,
\label{solF}}
where $μ$ can be evaluated directly from \eqref{solmu}.

It seems like we did not need to use equation \eqref{Fextn} in finding the solution to this extremization problem, which is of course not true. The Lagrange multiplier $μ$ depends on $A$ and $B$ so the solution for $η$ is essential and it follows by substituting \eqref{FextNL} and \eqref{Fext5} in \eqref{Fextn} (we make the $η$-dependence explicit now):
\equ{-\frac{A'(η)}{A(η)}\frac{n}{2n-1}μ +\frac{B'(η)}{B(η)}\frac{1}{2n-1}μ =0 \quad ⇒\quad \boxed{A(η)B'(η) -nB(η)A'(η)=0}\,.
\label{soln}}
Thus, solving this equation gives a solution for $η$.

Let us work out one simple example: Seiberg theory\cite{Seiberg:1996bd}, which consists of a single $\N=1$ vector multiplet with gauge group $USp(2N)$ coupled to one hypermultiplet in antisymmetric representation and $N_f<8$ hypermultiplets in fundamental rep. Putting this theory on $S^5$, the resulting matrix model for the free energy\cite{Jafferis:2012iv} is such that $n=3$ and the coefficients $A=\frac{9π}{8}$, $B=\frac{(8-N_f)π}{3}$ are constants. The equations \eqref{Fextn} and \eqref{soln} are then redundant and one can easily verify the well-known solution using \eqref{solmu} and \eqref{solF} with $\bar{F}=F_{S^5}N^{-\frac{5}{2}}$:
\equ{μ^2=\frac{4^4A^3}{3^3B}\qquad ⇒\qquad F_{S^5}^{Seiberg} =-N^{\frac{5}{2}}\frac{2}{5}\frac{9π}{\sqrt{2(8-N_f)}}\,·
\label{FS5Sbg}}

\section{\texorpdfstring{$\bm{(S^2_{ε_1}×S^1)×Σ_{\fg_2}}$}{(S²(ε₁)×S¹)×Σg₂}}\label{sec:5dUofSCI}
We consider now the Euclidean path integral on $(S^2_{ε_1}\times S^1_{r})×Σ_{\fg_2}$, where $ε_{1}$ is the refinement corresponding to the angular momentum along $S^2$ and $r$ is the radius of the $S^{1}$, which will be set to 1 shortly. The path integral on this background can be evaluated by supersymmetric localization, which localizes it on the north and south pole of the two $S^2$'s (for $\fg=0$; the generalization to $\fg≠0$ is obtained by invoking the A-twist formalism), where the space locally looks like the 5d $\Omega$-deformed background with parameters $\pm ε_{1,2}$ and we take $ε_2→0$. The partition function is thus obtained by gluing four copies of 5d Nekrasov partition functions given in Section \ref{sec:NPFs} with parameters identified as follows:
\equ{x^{(\ell)}=\begin{cases} e^{2\pi i r\(\tilde{u}+\frac{\fm_1ε_1}{2}+\frac{\fm_2ε_2}{2}\)} \\ e^{2\pi i r\(\tilde{u}+\frac{\fm_1ε_1}{2}-\frac{\fm_2ε_2}{2}\)} \\ e^{2\pi i r\(-\tilde{u}+\frac{\fm_1ε_1}{2}+\frac{\fm_2ε_2}{2}\)} \\ e^{2πi r\(-\tilde{u}+\frac{\fm_1ε_1}{2}-\frac{\fm_2ε_2}{2}\)} \end{cases}
\fq_{1}^{(\ell)}=\begin{cases} e^{2\pi i r ε_1} ≡\fq_1 \\ e^{2\pi i r ε_1} \\ e^{-2\pi i r ε_1} \\ e^{-2πi r ε_1} \end{cases}
\fq_{2}^{(\ell)}=\begin{cases} e^{2\pi i r ε_2} ≡\fq_2 \quad & \ell=nn \\ e^{-2\pi i r ε_2} \quad & \ell=ns \\ e^{2\pi i r ε_2} \quad & \ell=sn \\ e^{-2\pi i r ε_2} \quad & \ell=ss \end{cases}\,,
}
with similar expressions for $y^{(\ell)}$ depending on $\tnu,\fn_1,\fn_2$. Note that if we focus on just the $\tu$ and $ε_1$ terms, we see the ``identity gluing'' $\big(\tu^{(n\bullet)}=-\tu^{(s\bullet)}$ and $ε_1^{(n\bullet)}=-ε_1^{(s\bullet)}\big)$ for 3d SCI at work ensuring there is no twist of the $S^2$ corresponding to $ε_1$ \cite{Beem:2012mb,Closset:2018ghr}. We will work with complexified fugacities $\tnu$ and restrict the real part of fugacities to the domains $0 \leq \Re\,\tnu <\frac{1}{r}$, and $0 \leq \Re\,ε_{1,2}<\frac{1}{r}\,·$ We shall assume that instanton contributions are suppressed at large $N$ and thus the dominant contribution is entirely from the perturbative sector. We always choose the chamber where $|\fq_{i}|<1 \Leftrightarrow \Im\, ε_i >0$ but take the $ε_2→0$ limit after gluing the building blocks appropriately.

\paragraph{Classical.} The classical contribution in the perturbative sector follows immediately:
\equ{\Z^{cl}_{(S^2_{ε_1}×S^1)×Σ_{\fg_2}} =\text{exp}\(\frac{32π^3r}{g^2}\frac{\Tr(\fm_2\tu)}{ε_1} +2πi k r \Tr(\fm_1\fm_2\tu)\).
}
The YM term is subleading for the 5d quiver theories we consider and thus can be ignored. The CS term can be present for $SU(N)$ gauge groups when $N≥3$ but we will mostly set $k=0$.

\paragraph{Vector.} Let us first glue two blocks for $nn$ and $ns$ fixed points to get (suppressing unnecessary symbols)
\begingroup
\allowdisplaybreaks
\eqs{\Z^{1-loop,vec}_{(nn),(ns)} &=\left[\Big(x^{\alpha}\fq_{1}^{1+\frac{\fm_1}{2}}\fq_{2}^{1+\frac{\fm_2}{2}};\fq_{1},\fq_{2}\Big)\Big(x^{-\alpha}\fq_1^{-\frac{\fm_1}{2}}\fq_2^{-\frac{\fm_2}{2}};\fq_{1},\fq_{2}\Big)\right]^{\frac{1}{2}} \nn
&\qquad\qquad ×\left[\Big(x^{\alpha}\fq_{1}^{1+\frac{\fm_1}{2}}\fq_{2}^{-1-\frac{\fm_2}{2}};\fq_{1},\fq_{2}^{-1}\Big)\Big(x^{-\alpha}\fq_1^{-\frac{\fm_1}{2}}\fq_2^{\frac{\fm_2}{2}};\fq_{1},\fq_{2}^{-1}\Big)\right]^{\frac{1}{2}} \nn
&=\left[\frac{\Big(x^{\alpha}\fq_{1}^{1+\frac{\fm_1}{2}}\fq_{2}^{1+\frac{\fm_2}{2}};\fq_{1},\fq_{2}\Big)\Big(x^{-\alpha}\fq_1^{-\frac{\fm_1}{2}}\fq_2^{-\frac{\fm_2}{2}};\fq_{1},\fq_{2}\Big)}{\Big(x^{\alpha}\fq_{1}^{1+\frac{\fm_1}{2}}\fq_{2}^{-\frac{\fm_2}{2}};\fq_{1},\fq_{2}\Big)\Big(x^{-\alpha}\fq_1^{-\frac{\fm_1}{2}}\fq_2^{1+\frac{\fm_2}{2}};\fq_{1},\fq_{2}\Big)}\right]^{\frac{1}{2}}.
}
\endgroup
From the definition of double-Pochhammer in \eqref{dPdef}, it is clear that there is going to be a huge cancellation above, which in the limit $\fq_2→1$ gives
\equ{\Z^{1-loop,vec}_{(nn),(ns)} =\left[\frac{\Big(x^{\alpha}\fq_{1}^{1+\frac{\fm_1}{2}};\fq_{1}\Big)}{\Big(x^{-\alpha}\fq_1^{-\frac{\fm_1}{2}};\fq_{1}\Big)}\right]^{-\frac{\fm_2+1}{2}}.
}
Similarly, gluing the $sn$ and $ss$ blocks gives (in the limit $\fq_2→1$)
\equ{\Z^{1-loop,vec}_{(sn),(ss)} =\left[\frac{\Big(x^{\alpha}\fq_1^{1-\frac{\fm_1}{2}};\fq_1\Big)}{\Big(x^{-\alpha}\fq_{1}^{\frac{\fm_1}{2}};\fq_1\Big)}\right]^{-\frac{\fm_2+1}{2}}.
}
Now we combine the two results above to get the complete 1-loop vector contribution using the definitions and identities for $q$-Pochhammer symbol (see Appendix \ref{app:Id}):
\begingroup
\allowdisplaybreaks
\eqs{\Z^{1-loop,vec}_{(S^2_{ε_1}×S^1)×S^2} &=\left[\frac{\Big(x^{-\alpha}\fq_1^{-\frac{\fm_1}{2}};\fq_{1}\Big)}{\Big(x^{\alpha}\fq_{1}^{1+\frac{\fm_1}{2}};\fq_{1}\Big)}·\frac{\Big(x^{-\alpha}\fq_1^{\frac{\fm_1}{2}};\fq_1\Big)}{\Big(x^{\alpha}\fq_{1}^{1-\frac{\fm_1}{2}};\fq_1\Big)}\right]^{\frac{\fm_2+1}{2}} \label{VS2S1SLstep} \\
&=\left[(-x^α\sqrt{\fq_1})^{\frac{\fm_1}{2}}\frac{\Big(x^{-\alpha}\fq_1^{-\frac{\fm_1}{2}};\fq_{1}\Big)}{\Big(x^{\alpha}\fq_{1}^{1-\frac{\fm_1}{2}};\fq_{1}\Big)}\right]^{\fm_2+1}.
\label{VS2S1L}}
\endgroup
For general $\fg_2$, we will have the exponent $\fm_2+1-\fg_2$ by invoking the A-twist formalism. The form of the expression in \eqref{VS2S1L} is useful for comparison to 3d SCI contribution but we postpone that till the computation of the hyper contribution. For now, we massage the expression in \eqref{VS2S1SLstep} to generate $q$-theta functions, $Θ(x;q)=(x;q)(qx^{-1};q)$, which will turn out to be useful in taking the large $N$ limit (see Appendix \ref{app:Id} for more details). Reinstating the product over roots, we get
\equ{\Z^{1-loop,vec}_{(S^2_{ε_1}×S^1)×Σ_{\fg_2}} =∏_α\left[\frac{\Big(x^{-\alpha}\fq_1^{-\frac{α(\fm_1)}{2}};\fq_{1}\Big)^2}{Θ\Big(x^{\alpha}\fq_{1}^{1+\frac{α(\fm_1)}{2}};\fq_{1}\Big)}·\frac{\Big(x^{-\alpha}\fq_1^{\frac{α(\fm_1)}{2}};\fq_1\Big)^2}{Θ\Big(x^{\alpha}\fq_{1}^{1-\frac{α(\fm_1)}{2}};\fq_1\Big)}\right]^{\frac{α(\fm_2)+1-\fg_2}{2}}.
\label{VS2S1lN}}
We now define a function which will serve as a building block for all the 1-loop contributions
\equ{Z_B(x;q)=\frac{(qx^{-1};q)}{\sqrt{Θ(x;q)}}\,,
\label{defZB}}
such that the vector contribution looks like gluing of two 3d holomorphic blocks as follows
\equ{\Z^{1-loop,vec}_{(S^2_{ε_1}×S^1)×Σ_{\fg_2}} =∏_α\left[Z_B\Big(x^{\alpha}\fq_{1}^{1+\frac{α(\fm_1)}{2}};\fq_{1}\Big)·Z_B\Big(x^{\alpha}\fq_{1}^{1-\frac{α(\fm_1)}{2}};\fq_1\Big)\right]^{α(\fm_2)+1-\fg_2}.
\label{VS2S1hbf}}

\paragraph{Hyper.} Similarly, we glue four copies of the Nekrasov partition functions for hypermultiplet and get the complete 1-loop contribution:
\begingroup
\allowdisplaybreaks
\eqs{\Z^{1-loop,hyp}_{(S^2_{ε_1}×S^1)×S^2} &=\left[\frac{\Big(-x^{-ρ}\fq_{1}^{\frac{1-\fm_1}{2}};\fq_{1}\Big)}{\Big(-x^ρ\fq_1^{\frac{1+\fm_1}{2}};\fq_{1}\Big)}·\frac{\Big(-x^{-ρ}\fq_1^{\frac{1+\fm_1}{2}};\fq_1\Big)}{\Big(-x^{ρ}\fq_{1}^{\frac{1-\fm_1}{2}};\fq_1\Big)}\right]^{-\frac{\fm_2}{2}} \nn
&=\left[(x^ρ)^{\frac{\fm_1}{2}}\frac{\Big(-x^{-ρ}\fq_1^{\frac{1-\fm_1}{2}};\fq_{1}\Big)}{\Big(-x^ρ\fq_{1}^{\frac{1-\fm_1}{2}};\fq_{1}\Big)}\right]^{-\fm_2}.
\label{HS2S1L}}
\endgroup
The expression inside the brackets matches the contribution of 3d chiral multiplet with R-charge $ν_r=1$ to the 3d SCI \cite{Hwang:2012jh} (up to the $-$ signs that may be absorbed in flavour fugacity and identify $\fm=-\fm_1$):
\equ{Z_{3d}(x\fq^{\nu_{r}/2},\fm;\fq)= \(\fq^{(1-\nu_{r})/2}x^{-1}\)^{\fm/2}\frac{(\fq^{1-\nu_{r}/2+\fm/2}x^{-1};\fq)}{(\fq^{\nu_{r}/2+\fm/2}x;\fq)}\,·
}
Similarly the vector contribution in \eqref{VS2S1L} can be seen to be the contribution of a chiral multiplet with $ν_r=2$, which follows from the usual argument (see, for example, \cite{Willett:2016adv}) about a massive vector not contributing to the index and a vector can be made massive by coupling it to a $ν_r=0$ chiral field. Then using the fact that $Z_{3d}^{chiral}(ν_r=0)·Z_{3d}^{chiral}(ν_r=2)=1$ proves the claim in the previous sentence.

Rewriting \eqref{HS2S1L} in terms of $Z_B(x;q)$ as for the vector, while reinstating the product over weights and fugacities, we get
\equ{\Z^{1-loop,hyp}_{(S^2_{ε_1}×S^1)×Σ_{\fg_2}} =∏_ρ\left[Z_B\Big(-x^{ρ}y\fq_{1}^{\frac{1+ρ(\fm_1)}{2}};\fq_{1}\Big)·Z_B\Big(-x^{ρ}y\fq_{1}^{\frac{1-ρ(\fm_1)}{2}};\fq_1\Big)\right]^{-ρ(\fm_2)-\hat{\fn}_2},
\label{HS2S1lN}}
where we will use $\hat{\fn}_2=\fn_2(1-\fg_2)$ for general $\fg_2$.

\paragraph{Complete Perturbative Result.} Combining all the above results, we get the perturbative partition function:
\equ{Z^{pert}_{(S^2_{ε_1}×S^1)×Σ_{\fg_2}}=\frac{1}{|W_G|}∑_{\fm_1,\fm_2}∮d\tu\,H^{\fg}\,e^{\frac{32π^3}{g^2}\frac{\Tr(\fm_2\tu)}{ε_1} +2πik \Tr(\fm_1\fm_2\tu)} \Z^{1-loop,vec}_{(S^2_{ε_1}×S^1)×Σ_{\fg_2}}\Z^{1-loop,hyp}_{(S^2_{ε_1}×S^1)×Σ_{\fg_2}}\,,
}
where $H$ is the Hessian appearing in the handle gluing operator $\H$. Comparing with \eqref{jkformula}, we can identify the following operators
\begingroup
\allowdisplaybreaks
\eqs{Π^{pert}_a &=e^{\frac{32π^3}{g^2}\frac{\tu^a}{ε_1} +2πik (\fm_1\tu)^a}∏_α\left[Z_B\Big(x^{\alpha}\fq_{1}^{1+\frac{α(\fm_1)}{2}};\fq_{1}\Big)·Z_B\Big(x^{\alpha}\fq_{1}^{1-\frac{α(\fm_1)}{2}};\fq_1\Big)\right]^{α^a}∏_ρ\left[⋯\right]^{-ρ^a} \label{Paop}\\
Π^{pert}_i &= ∏_ρ\left[Z_B\Big(-x^{ρ}y\fq_{1}^{\frac{1+ρ(\fm_1)}{2}};\fq_{1}\Big)·Z_B\Big(-x^{ρ}y\fq_{1}^{\frac{1-ρ(\fm_1)}{2}};\fq_1\Big)\right]^{-1} \label{Piop}\\
\H^{pert} &=∏_α\left[Z_B\Big(x^{\alpha}\fq_{1}^{1+\frac{α(\fm_1)}{2}};\fq_{1}\Big)·Z_B\Big(x^{\alpha}\fq_{1}^{1-\frac{α(\fm_1)}{2}};\fq_1\Big)\right]^{-1}\det_{ab}\frac{∂\log(Π^{pert}_a)}{2πi ∂\tu_b} \nn
&=∏_{α>0}\left[(-1)^{α(\fm_1)}\fq_1^{-\frac{α(\fm_1)}{2}}\Big(\fq_1^{\frac{α(\fm_1)}{2}}x^α -1\Big)\Big(\fq_1^{\frac{α(\fm_1)}{2}}x^{-α}-1\Big)\right]^{-1}\det_{ab}\frac{∂\log(Π^{pert}_a)}{2πi ∂\tu_b}\,·
\label{Hop}}
\endgroup
The $⋯$ in \eqref{Paop} is same as the product of two $Z_B$-factors appearing in \eqref{Piop}. The flavour magnetic fluxes $\fn_1$ are not explicitly written above, but they can be introduced by shifting $y→y\fq_1^{\frac{\fn_1}{2}}$ whenever needed. Of course, the perturbative Bethe vacua follows from
\equ{\S^{pert}_{BE}=\left\{\hat{u}\;|\;Π^{pert}_a(\hat{u})≡\text{exp}\(2πi\tfrac{∂\W^{pert}}{∂\tu_a}(\hat{u})\)=1,\quad a=1,⋯,r_G\right\}/W_G,
\label{BetheEq3dSCI}}
from where we find the perturbative effective twisted potential $\W^{pert}_{(S^2_{ε_1}×S^1)×Σ_{\fg_2}}$. The classical contribution simply reads
\equ{\W^{cl}_{(S^2_{ε_1}×S^1)×Σ_{\fg_2}}=\frac{8π^2\tu^2}{i g^2 ε_1} +\frac{k}{2}\Tr(\fm_1\tu^2)\,.
}
The 1-loop contributions to $\W$ involve integrating the function $\log Z_B$. This is quite cumbersome in general but since we are interested in the large $N$ limit mostly, we can write $\log Z_B(x;q)$ always in terms of $\log Θ(x;q)$ by making sure that the $\log(qx^{-1};q)→0$ as $x→∞$. The large $N$ expansion of $\log Θ(x;q)$ is discussed in the Appendix \ref{app:Id} and we can define the associated integrated function using \eqref{logThetaId} as follows:
\eqs{∫dz \log Θ\(z;ω\)^{-1} &≈∫dz \left[πi T\(z|ω\)-\half\right] \nn
&=\frac{πi z^3}{3ω} -\frac{πi (ω-1)z^2}{2ω} +\left[\frac{πi (ω^2+3ω-1)}{6ω} -\frac{1}{2}\right]z ≡\check{T}(z|ω)\,. \nn
\text{Also, } ∫dz \log Θ\(z+ζ;ω\)^{-1} &≈\check{T}(z+ζ|ω)-\check{T}(ζ|ω)≡\check{T}(z,ζ|ω)\,.
\label{defTcheck}}
Using these integrals, the vector and hyper contributions to $\W$ read
\begingroup
\allowdisplaybreaks
\eqs{(2πi)\W^{1-loop,vec}_{(S^2_{ε_1}×S^1)×Σ_{\fg_2}} &=\frac{1}{2}∑_{α>0,±}\left[\check{T}\(α(\tu),1+ε_1±\tfrac{ε_1α(\fm_1)}{2}|ε_1\) +\check{T}\(α(\tu),±\tfrac{ε_1α(\fm_1)}{2}|ε_1\)\right], \\
(2πi)\W^{1-loop,hyp}_{(S^2_{ε_1}×S^1)×Σ_{\fg_2}} &=-\frac{1}{2}∑_{ρ}\left[\check{T}\(ρ(\tu),\tnu+\tfrac{1+ε_1+ε_1ρ(\fm_1)}{2}|ε_1\) +\check{T}\(ρ(\tu),\tnu+\tfrac{1+ε_1-ε_1ρ(\fm_1)}{2}|ε_1\)\right].
}
\endgroup
Note that we have assumed $ρ>0$ to write the above form of the hyper contributions. For explicit representations, one has to make sure that this is true by switching between the analogues of the two terms present in the vector contribution.

After we find the Bethe vacua solutions from extremizing the twisted superpotential at large $N$, we can use this solution to write the final result for the partition function at large $N$ as follows:
\equ{Z_{(S^2_{ε_1}×S^1)×\Sigma_{\fg_2}}(\tnu;\fn_1,\fn_2) ≈\sum_{\hat{u} \in \S_{\mathit{BE}}} \Pi_i(\hat{u},\nu)^{\fn_{2i}(1-\fg_2)} \H(\hat{u},\nu)^{\fg_2-1}\,,
\label{ZgenPHform}}
with the relevant operators explicitly given in \eqref{Piop} and \eqref{Hop}. As explained before, the instanton contributions are subleading at large $N$ so we drop the specifier `\emph{pert}' above.

We can rewrite the above expression for the partition function at large $N$ in terms of the twisted superpotential as was done for the the partition function on $S^3_b×Σ_{\fg}$ in \cite{Crichigno:2020ouj}. We ignore the subleading Hessian contribution and then, schematically, \eqref{ZgenPHform} takes the following form
\eqs{\log Z_{(S^2_{ε_1}×S^1)×\Sigma_{\fg_2}} &≈-∑_{α∈Ad(G)'}(\fg_2-1)\tfrac{πi}{2}T(α(\hat{u})|ε_1) +∑_I∑_{ρ∈R_I}\fn_{2I}(\fg_2-1)\tfrac{πi}{2}T(ρ(\hat{u})+\tnu_I|ε_1) \nn
&≈∑_I∑_{ρ∈R_I}(\fg_2-1)\(\fn_{2I}-\frac{1+ε_1}{2\tnu_I}\)\frac{∂\big[\frac{1}{2}\check{T}(ρ(\tu),\tnu_I|ε_1)\big]}{∂\tu}(\hat{u}) \nn
&≈-2πi(\fg_2-1)∑_I\(\fn_{2I}-\frac{1+ε_1}{2\tnu_I}\)\frac{∂\W_{(S^2_{ε_1}×S^1)×Σ_{\fg_2}}(\tnu)}{∂\tnu_I}\,,
\label{logZgennuinv}}
where the second line follows from the large $N$ limit of the Bethe equations in \eqref{BetheEq3dSCI} that allows one to replace the sum over vectors by a sum over hypermultiplets, and the last line follows from identifying $\frac{\check{T}}{2}$ as $(2πi)$ times the twisted superpotential. The switching of the derivative of $\W$ from $\tu$ to $\tnu$ introduces an overall minus sign due to the definition \eqref{defTcheck}. This is as far as we can go in general, but for the quiver theories with massive type IIA string duals, the above expression can be cast into a more useful form \eqref{logZ3dSCIrelWgen}. We do that now by evaluating the $Π$'s and $\H$'s explicitly at large $N$ for such theories.

\subsection[\texorpdfstring{Large $N$ Limit (mIIA Duals)}{Large N Limit (mIIA Duals)}]{Large $\bm{N}$ Limit (mIIA Duals)}
We note that in the 1-loop determinants, only two combinations of the gauge parameter and flux, $\tu±\frac{1}{2}ε_1\fm_1$, appear, so we choose an ansatz with both $\tu$ and $\fm_1$ having the same large $N$ behaviour. Explicitly,
\equ{\tu^a =-iN^{α}x\,,\quad \fm_1^a =-iN^{α} η x\qquad ⇒\quad ∑_a →N∫dx ρ(x)\,,
}
where $η$ needs to be determined and we introduced the eigenvalue density $ρ(x)$ satisfying $∫_0^{x^*}ρ(x)dx=1$. The explicit sums over roots and weights for vectors and hypermultiplets in given representations, respectively, are well-known by now so we do not spell them out here. Readers new to this computation are urged to browse \cite[⋯]{Jafferis:2012iv,Crichigno:2018adf} for relevant details before proceeding further.

\paragraph{Seiberg Theory.} Let us work out this simplest example, which can then be generalized for generic quiver theories following \cite{Crichigno:2018adf}. First, we deal with the twisted superpotential and present the \emph{north} and \emph{south} blocks separately (to motivate a few definitions that will follow later):
\begingroup
\allowdisplaybreaks
\eqs{(2πi)\W^{Seiberg}_{(S^2_{ε_1}×S^1)×Σ_{\fg_2}} &=-\frac{π}{4}N^{\frac{5}{2}}\left[\frac{(8-N_f)(4+3ε_1 η(2+ε_1η))}{3ε_1}∫ρ(x)x^3dx\right. \nn
&\qquad \left. -\frac{(1+ε_1)^2-(2\tnu +ε_1\fn_1)^2}{2ε_1}∫dxdyρ(x)ρ(y)(|x+y|+|x-y|)\right] \nn
&\quad -\frac{π}{4}N^{\frac{5}{2}}\left[\frac{(8-N_f)(4+3ε_1 η(-2+ε_1η))}{3ε_1}∫ρ(x)x^3dx\right. \nn
&\qquad \left. -\frac{(1+ε_1)^2-(2\tnu -ε_1\fn_1)^2}{2ε_1}∫dxdyρ(x)ρ(y)(|x+y|+|x-y|)\right] \nn
&=-\frac{π}{2}N^{\frac{5}{2}}\left[\frac{(8-N_f)(4+3ε_1^2η^2)}{3ε_1}∫ρ(x)x^3dx\right. \nn
&\qquad \left. -\frac{(1+ε_1)^2 -ε_1^2\fn_1^2-4\tnu^2}{2ε_1}∫dxdyρ(x)ρ(y)|x±y|\right],
\label{W3dSCIintg}}
\endgroup
where we have used the notation $|x±y|$ to denote a sum of two terms in the last line. We have already used the fact that the scaling behaviour of the two integrals demand $N^{1+3α}=N^{2+α}$, which gives $α=\frac{1}{2}$ leading to the expected $N^{\frac{5}{2}}$ scaling for $\W$ above. Now we have to extremize the above function with respect to both $ρ(x)$ and $η$ and find the appropriate solution. Comparing \eqref{W3dSCIintg} with \eqref{Fdef}, we find $\tilde{F}=-(2πi)\W\frac{2}{π}N^{-\frac{5}{2}}$ with $n=3$, $A(η)=\frac{(1+ε_1)^2 -ε_1^2\fn_1^2-4\tnu^2}{2ε_1}$, $B(η)=\frac{(8-N_f)(4+3ε_1^2η^2)}{3ε_1}$. Then, \eqref{soln} leads to $η=0$ and using \eqref{solmu} in \eqref{solF}, gives
\eqs{(2πi)\W^{Seiberg}_{(S^2_{ε_1}×S^1)×Σ_{\fg_2}} &=-\(-\frac{π}{2}N^{\frac{5}{2}}\)\frac{2}{5}\frac{2\sqrt{2}\((1+ε_1)^2 -ε_1^2\fn_1^2-4\tnu^2\)^{\frac{3}{2}}}{3ε_1\sqrt{8-N_f}} \nn
&=-\frac{2}{27ε_1}\((1+ε_1)^2 -ε_1^2\fn_1^2-4\tnu^2\)^{\frac{3}{2}}F^{Seiberg}_{S^5}
}
where the second line follows after using \eqref{FS5Sbg}. It is clear that this solution does not admit factorization for finite $ε_1$. Defining
\equ{\tnu^{±}=\frac{2\tnu±ε_1\fn_1}{1+ε_1}\quad \text{ and }\quad b=\sqrt{ε_1},
}
we can relate the twisted superpotential to the mass-deformed, squashed 5-sphere partition function, defined in \eqref{FS5mexp}:
\equ{(2πi)\W^{Seiberg}_{(S^2_{ε_1}×S^1)×Σ_{\fg_2}} =-\frac{8(1+ε_1)}{27}F_{S^5_{(b,b^{-1},Q)}}\(\tfrac{3Q\tnu^{±}}{2}\),
\label{W3dSCIsbg}}
where $Q=\frac{1}{2}(b+b^{-1})$. The notation $F_{S^5_{\vec{ω}}}(m^{±})$ above just means that there are two mass parameters and the numerator in \eqref{FS5mexp} is to be evaluated accordingly. So for the above expression, it would read $[(1-Q^2(\tnu^+)^2)+(1-Q^2(\tnu^-)^2)]^{\frac{3}{2}}$, which one can verify gives the prefactor in \eqref{FS5Sbg}. In general, the summation $∑_I$ in \eqref{FS5mexp} becomes $∑_{I,±}$ and $c^I_ρ$'s corresponding to $\tnu_I$ are used for both $\tnu^{±}_I$. It is worthwhile to mention here that once we set $η=0$ in \eqref{W3dSCIintg}, the resulting integral expression already satisfies \eqref{W3dSCIsbg}, so \eqref{W3dSCIsbg} is valid both ``off-shell'' and ``on-shell''. 

Now, we move on to the partition function, which follows from \eqref{ZgenPHform} or \eqref{logZgennuinv}:
\begingroup
\allowdisplaybreaks
\eqs{\log Z_{(S^2_{ε_1}×S^1)×Σ_{\fg_2}}^{Seiberg} &=-πN^{\frac{5}{2}}(\fg_2-1)\frac{\big[\frac{1+ε_1}{2}-\fn_2\(\tnu +\frac{ε_1\fn_1}{2}\)\big]+\big[\frac{1+ε_1}{2}-\fn_2\(\tnu -\frac{ε_1\fn_1}{2}\)\big]}{ε_1} \nn
&\qquad ×∫dxdyρ(x)ρ(y)|x±y| \nn
&=\frac{8}{9}(\fg_2-1)\frac{(1+ε_1 -2\fn_2\tnu)}{2ε_1}\sqrt{(1+ε_1)^2-ε_1^2\fn_1^2-4\tnu^2}\,F_{S^5}^{Seiberg} \label{logZSbgexp}\\
⇒\log Z_{(S^2_{ε_1}×S^1)×Σ_{\fg_2}}^{Seiberg} &=2\log Z_{S^3_{\sqrt{ε_1}}×Σ_{\fg_2}}\(\tnu^{±};\fn_2\).
\label{FullCardyF}}
\endgroup
It is again obvious from \eqref{logZSbgexp} that factorization is not possible for finite $ε_1$. The coefficient of the integral expression is split in such a way that leads to factorization expected of the 3d Cardy formula \cite{Choi:2019dfu} only in the strict Cardy limit $(ε_1≪1)$. The simple relation in \eqref{FullCardyF} follows from \cite{Crichigno:2018adf},\footnote{For reference, we reproduce here the expression for $S^3_b×Σ_{\fg}$ partition function of general quivers:
\[\log Z_{S^3_b×Σ_{\fg}}(\tnu;\fn)=\frac{8}{9}Q^2(\fg-1)\frac{\left(∑_Ic_ρ^I(1-\fn_I\tnu_I)\right)\left(∑_Ic_ρ^I(1-\tnu_I^2)\right)^{\frac{1}{2}}}{\left(∑_Ic_ρ^I\right)^{\frac{3}{2}}}F_{S^5}\,.\]} and is similar to relation 2 in Table \ref{tab:relPFs} between 5d SCI and squashed $S^5$ free energy\cite{Crichigno:2020ouj}. The difference being that the quivers are not the same on both sides of the equality and the doubling of chemical potential $\tnu→\tnu^{±}$ seems formal rather than implying a deep relation between different quiver theories.\footnote{Though, it is curious that the $(S^2_{ε_1}×S^1)×Σ_{\fg_2}$ partition function of Seiberg theory at large $N$ is given by the square of the $S^3_{\sqrt{ε_1}}×Σ_{\fg_2}$ partition function of class (C) orbifold theory with $k=1$ \cite[Sec. 3.1]{Crichigno:2018adf}, with parameters identified as in \eqref{FullCardyF}.}

\paragraph{Generic Quivers.} Following the generalization discussed in \cite{Crichigno:2018adf} to a family of generic quivers with behaviour similar to the Seiberg theory, we can generalize above results to such theories here too. We simply state the general results as the matrix model analysis is a straightforward extension of the Seiberg theory:
\begingroup
\allowdisplaybreaks
\eqsc{(2πi)\W_{(S^2_{ε_1}×S^1)×Σ_{\fg_2}}(\tnu_I;\fn_{1I}) =-\frac{8(1+ε_1)}{27}F_{S^5_{(b,b^{-1},Q)}}\Big(\tfrac{3Q\tnu^{±}_I}{2}\Big) \label{W3dSCIgen}\\
\log Z_{(S^2_{ε_1}×S^1)×Σ_{\fg_2}}(\tnu_I;\fn_{1I},\fn_{2I}) =2\log Z_{S^3_{\sqrt{ε_1}}×Σ_{\fg_2}}\(\tnu^{±}_I;\fn_{2I}\) \label{logZ3dSCIrelS3gen}\\
\log Z_{(S^2_{ε_1}×S^1)×Σ_{\fg_2}}(\tnu_I;\fn_{1I},\fn_{2I}) =-2πi(\fg_2-1)\bigg[4\W +∑_I\(\fn_{2I} -2\tnu_I\)\frac{∂\W}{∂\tnu_I} -2ε_1\frac{∂\W}{∂ε_1}\bigg].
\label{logZ3dSCIrelWgen}}
\endgroup
The last relation follows from \eqref{logZgennuinv} by verifying that the following relation holds for the $\W$ appearing above:
\equ{4\W_{(S^2_{ε_1}×S^1)×Σ_{\fg_2}}=∑_I\(2\tnu_I -\frac{1+ε_1}{2\tnu}\)\frac{∂\W}{∂\tnu_I} +2ε_1\frac{∂\W}{∂ε_1}\,·
}
If we use $\tnu^{±}_I$ instead of $\tnu_I$, we get another relation that directly verifies \eqref{W3dSCIgen} and \eqref{logZ3dSCIrelS3gen} \cite[Sec. 5]{Crichigno:2020ouj}:
\equ{\log Z_{(S^2_{ε_1}×S^1)×Σ_{\fg_2}} =-(2πi)\frac{(\fg_2-1)}{(1+ε_1)}\bigg[6\W +2∑_{±,I}\(\fn_{2I} -\tnu^{±}_I\)\frac{∂\W}{∂\tnu^{±}_I}\bigg].
\label{logZ3dSCIrelWS3gen}}

\

We can compactify the above expressions even more by choosing a quartet of constrained flavour fugacities and fluxes for each parameter in the trio $\{\tnu_I,\fn_{1I},\fn_{2I}\}$ as follows:
\begingroup
\allowdisplaybreaks
\eqsc{\tnu^{+±}_I =(1+ε_1)±(2\tnu_I +ε_1\fn_{1I})\,,\quad \tnu^{-±}_I =(1+ε_1)±(2\tnu_I -ε_1\fn_{1I})\,, \\
\fn_{iI}^{+±}=2(1±\fn_{iI})\,,\quad \fn_{iI}^{-±}=2(1±\fn_{iI})\quad\text{ for }i=1,2\,. \\
(2πi)\W_{(S^2_{ε_1}×S^1)×Σ_{\fg_2}} =-\frac{2}{27 ε_1}\(\frac{{\textstyle ∑_I}c_{ρ}^I\big[\tnu_I^{++}\tnu_I^{+-} +\tnu_I^{-+}\tnu_I^{--}\big]}{2∑_Ic_{ρ}^I}\)^{\frac{3}{2}}F_{S^5}\label{W3dSCIFS5} \\
\log Z_{(S^2_{ε_1}×S^1)×Σ_{\fg_2}} =-2πi(\fg_2-1)∑_{I,\{±\}}\fn_{2I}^{±±}\frac{∂\W}{∂\tnu_I^{±±}}\,·
\label{logZ3dSCIrelWgenComp}}
\endgroup
Note that \eqref{W3dSCIFS5} does not have $∑_{I,±}$ as both sides of the equation refer to the same quiver theory, in contrast to the form present in \eqref{W3dSCIsbg} or \eqref{logZ3dSCIrelS3gen}.

\paragraph{Universal.} Now let us set all the flavour chemical potentials and fluxes to zero and analyze the resulting universal behaviour of the partition function. The following relations then hold for generic quivers:
\eqsg{(2πi)\W^{univ}_{(S^2_{ε_1}×S^1)×Σ_{\fg_2}}=-\frac{2}{27}\frac{(1+ε_1)^3}{ε_1}F_{S^5} \\
\log Z^{univ}_{(S^2_{ε_1}×S^1)×Σ_{\fg_2}}=2\log Z_{S^3_{\sqrt{ε_1}}×Σ_{\fg_2}} =\frac{(1+ε_1)^2}{2ε_1}\log Z_{S^3×Σ_{\fg}}\,.
}
The latter relation can be brought to a familiar form by substituting $ε_1=\frac{ω}{2πi}$ and using the supersymmetric constraint $ω+2πi=2ϕ$ from \cite{Bobev:2019zmz}:
\equ{\log Z^{univ}_{(S^2_{ω}×S^1)×Σ_{\fg_2}} =\frac{ϕ^2}{πiω}\log Z^{univ}_{S^3×Σ_{\fg_2}}\,.
}
This verifies the holographic prediction for 5d SCFTs dual to rotating, charged $AdS_6$ black holes.

\paragraph{Extremization.} An explicit form for the partition function useful for extremization with respect to $\tnu$ can be written down following \cite{Crichigno:2018adf} leading to
\equ{\log Z_{(S^2_{ε_1}×S^1)×Σ_{\fg_2}} =\frac{\Big((1+ε_1)∑_Ac_{α}^A -2∑_Ic_{ρ}^I\fn_{2I}\tnu_I\Big)\Big(∑_Ic_{ρ}^I\big[(1+ε_1)^2\hat{\fn}_1^2 -4\tnu_I^2\big]\Big)^{\frac{1}{2}}}{2ε_1\big(∑_Ic_{ρ}^I\big)^{\frac{3}{2}}}\log Z_{S^3×Σ_{\fg_2}},
}
where we have defined
\equ{\hat{\fn}_1^2= \Big({\textstyle ∑_Ic_{ρ}^I}\Big)^{-1}{\textstyle ∑_Ic_{ρ}^I}\Big(1-\tfrac{ε_1^2\fn_{1I}^2}{(1+ε_1)^2}\Big)\,,\quad \hat{\fn}_2^2=\Big({\textstyle ∑_Ic_{ρ}^I}\Big)^{-1}{\textstyle ∑_Ic_{ρ}^I}\fn_{2I}^2 \,.
}
Now, extremization leads to the following result for the partition function:
\begingroup
\allowdisplaybreaks
\eqsc{∂_{\tnu_I}\log Z =0 \quad ⇒ \quad \tnu^{(±)}_I =\frac{(1+ε_1)\fn_{2I}}{8\hat{\fn}_2^2}\bigg(1±\sqrt{1+8\hat{\fn}_1^2\hat{\fn}_2^2}\bigg) \\
⇒\log Z_{(S^2_{ε_1}×S^1)×Σ_{\fg_2}} =\frac{(1+ε_1)^2}{2ε_1}\(±\tfrac{\sqrt{2\hat{\fn}_1^2}\,\big|\hat{\fn}_1^2\hat{\fn}_2^2 -1\big|^{\frac{3}{2}}\big(\sqrt{1+8\hat{\fn}_1^2\hat{\fn}_2^2}±1\big)}{\big|4\hat{\fn}_1^2\hat{\fn}_2^2 -1 ∓\sqrt{1+8\hat{\fn}_1^2\hat{\fn}_2^2}\big|^{\frac{3}{2}}}\)\log Z_{S^3×Σ_{\fg_2}}\,.
}
\endgroup
This should be useful in performing nontrivial and non-universal holographic checks, similar to \cite{Bah:2018lyv}.

\section{\texorpdfstring{$\bm{(\wt{S^2_{ε_1}}×S^1)×Σ_{\fg_2}}$}{(S̃²(ε₁)×S¹)×Σg₂}}\label{sec:5dUofTTI}
We consider here the partition function on $\wt{S^2_{ε_1}}×S^1$, with $\wt{S^2}$ denoting a partial topological twist on $S^2$. The unrefined ($ε_1=0$) version of this 5d index has been discussed before in \cite{Crichigno:2018adf,Hosseini:2018uzp} so some similarity in the results will be inevitable. But we will construct various contributions to the refined version of this 5d partition function by gluing the building blocks \eqref{5dblock} given in Section \ref{sec:NPFs}, as done in the previous section. The identification of the four sets of parameters at four fixed points is as follows:
\equ{x^{(\ell)}=\begin{cases} e^{2\pi i r\(\tilde{u}+\frac{\fm_1ε_1}{2}+\frac{\fm_2ε_2}{2}\)} \\ e^{2\pi i r\(\tilde{u}+\frac{\fm_1ε_1}{2}-\frac{\fm_2ε_2}{2}\)} \\ e^{2\pi i r\(\tilde{u}-\frac{\fm_1ε_1}{2}+\frac{\fm_2ε_2}{2}\)} \\ e^{2πi r\(\tilde{u}-\frac{\fm_1ε_1}{2}-\frac{\fm_2ε_2}{2}\)} \end{cases}
\fq_{1}^{(\ell)}=\begin{cases} e^{2\pi i r ε_1} ≡\fq_1 \\ e^{2\pi i r ε_1} \\ e^{-2\pi i r ε_1} \\ e^{-2πi r ε_1} \end{cases}
\fq_{2}^{(\ell)}=\begin{cases} e^{2\pi i r ε_2} ≡\fq_2 \quad & \ell=nn \\ e^{-2\pi i r ε_2} \quad & \ell=ns \\ e^{2\pi i r ε_2} \quad & \ell=sn \\ e^{-2\pi i r ε_2} \quad & \ell=ss \end{cases}\,,
}
We will work with complexified fugacities as before. Note that if we focus on just the $\tu$ and $ε_1$ terms, we see another ``identity gluing'' $\big(\tu^{(n\bullet)}=\tu^{(s\bullet)}$ and $ε_1^{(n\bullet)}=-ε_1^{(s\bullet)}\big)$ valid for 3d TTI at work, which now ensures there is a twist of the $S^2$ corresponding to $ε_1$ \cite{Closset:2018ghr}. We shall again assume that instanton contributions are suppressed at large $N$ and thus the dominant contribution is entirely from the perturbative sector.

\paragraph{Classical.} The classical contribution again follows immediately and matches the unrefined result:
\equ{\Z^{cl}_{(\wt{S^2_{ε_1}}×S^1)×S^2} =\text{exp}\(\frac{16π^3r}{g^2}\Tr(\fm_1\fm_2) +2πi k r \Tr(\fm_1\fm_2\tu)\).
}
As in the previous section, we will set $k=0$ and ignore the subleading contribution of the YM term in the large $N$ analysis.

\paragraph{Vector.} Let us now glue a pair of blocks $nn$ and $ns$ to get in the limit $\fq_2→1$ (suppressing unnecessary symbols)
\equ{\Z^{1-loop,vec}_{(nn),(ns)} =\left[\frac{\Big(x^{\alpha}\fq_{1}^{1+\frac{\fm_1}{2}};\fq_{1}\Big)}{\Big(x^{-\alpha}\fq_1^{-\frac{\fm_1}{2}};\fq_{1}\Big)}\right]^{-\frac{\fm_2+1}{2}}.
}
Similarly, gluing the $sn$ and $ss$ blocks gives in the limit $\fq_2→1$
\equ{\Z^{1-loop,vec}_{(sn),(ss)} =\left[\frac{\Big(x^{\alpha}\fq_1^{-\frac{\fm_1}{2}};\fq_1\Big)}{\Big(x^{-\alpha}\fq_{1}^{1+\frac{\fm_1}{2}};\fq_1\Big)}\right]^{\frac{\fm_2+1}{2}}.
}
Now the two results above can be combined to get the complete 1-loop vector contribution (see Appendix \ref{app:Id} for relevant identities):
\begingroup
\allowdisplaybreaks
\eqs{\Z^{1-loop,vec}_{(\wt{S^2_{ε_1}}×S^1)×S^2} &=\left[\frac{\Big(x^{-\alpha}\fq_1^{-\frac{\fm_1}{2}};\fq_{1}\Big)}{\Big(x^{\alpha}\fq_{1}^{1+\frac{\fm_1}{2}};\fq_{1}\Big)}·\frac{\Big(x^{\alpha}\fq_1^{-\frac{\fm_1}{2}};\fq_1\Big)}{\Big(x^{-\alpha}\fq_{1}^{1+\frac{\fm_1}{2}};\fq_1\Big)}\right]^{\frac{\fm_2+1}{2}} \label{VS2S1tSLstep}\\
&=\left[(-x^{α})^{\frac{\fm_1+1}{2}}\Big(x^{-α}\fq_1^{\frac{1-|\fm_1+1|}{2}};\fq_1\Big)_{|\fm_1+1|}^{\sgn(\fm_1+1)}\right]^{\fm_2+1}.
\label{VS2S1tL}}
\endgroup
For general $\fg_2$, we will have the exponent $\fm_2+1-\fg_2$. Note that in the unrefined case, $ε_1=0\,⇒\,\fq_1=1$ and the finite $\fq_1$-Pochhammer symbol reduces to just $(1-x^{-α})^{\fm_1+1}$, which matches the expressions derived in \cite{Crichigno:2018adf,Hosseini:2018uzp}. In addition, the expression in \eqref{VS2S1tL} is also useful for comparing to vector contribution to 3d refined TTI in \cite{Benini:2015noa}, but it does not yield nicely to the large $N$ limit. So, as done in the previous section, we rewrite the expression in \eqref{VS2S1tSLstep} using the function $Z_B(x;q)$ defined in \eqref{defZB}, which has already proven useful while taking the large $N$ limit. Reinstating the product over roots, we get
\equ{\Z^{1-loop,vec}_{(\wt{S^2_{ε_1}}×S^1)×Σ_{\fg_2}} =∏_{α}\left[Z_B\Big(x^{\alpha}\fq_{1}^{1+\frac{α(\fm_1)}{2}};\fq_1\Big)·Z_B\Big(x^{\alpha}\fq_{1}^{-\frac{α(\fm_1)}{2}};\fq_1\Big)^{-1}\right]^{α(\fm_2)+1-\fg_2}.
}

\paragraph{Hyper.} Similarly, gluing four copies of the Nekrasov partition functions for hypermultiplet gives us the following 1-loop contribution:
\eqs{\Z^{1-loop,hyp}_{(\wt{S^2_{ε_1}}×S^1)×S^2} &=\left[\frac{\Big(-x^{-ρ}\fq_{1}^{\frac{1-\fm_1}{2}};\fq_{1}\Big)}{\Big(-x^ρ\fq_1^{\frac{1+\fm_1}{2}};\fq_{1}\Big)}·\frac{\Big(-x^{ρ}\fq_1^{\frac{1-\fm_1}{2}};\fq_1\Big)}{\Big(-x^{-ρ}\fq_{1}^{\frac{1+\fm_1}{2}};\fq_1\Big)}\right]^{-\frac{\fm_2}{2}} \nn
&=\left[(x^{ρ})^{\frac{\fm_1}{2}}\Big(-x^{-ρ}\fq_1^{\frac{1-|\fm_1|}{2}};\fq_1\Big)_{|\fm_1|}^{\sgn(\fm_1)}\right]^{-\fm_2}.
\label{HS2S1tL}}
Again, the above result compares favourably to the chiral multiplet's contribution to 3d refined TTI (up to the $-$ sign which may be absorbed in the flavour fugacity) \cite[Sec. 4]{Benini:2015noa}. Rewriting the above result in terms of $Z_B(x;q)$ and reinstating the product over weights and fugacities, we get
\equ{\Z^{1-loop,hyp}_{(\wt{S^2_{ε_1}}×S^1)×Σ_{\fg_2}} =∏_{ρ}\left[Z_B\Big(-x^ρy\fq_1^{\frac{1+ρ(\fm_1)}{2}};\fq_1\Big)·Z_B\Big(-x^{ρ}y\fq_1^{\frac{1-ρ(\fm_1)}{2}};\fq_1\Big)^{-1}\right]^{-ρ(\fm_2)-\hat{\fn}_2},
}
where again we use $\hat{\fn}_2=\fn_2(1-\fg_2)$ for general $\fg_2$.

\paragraph{Complete Perturbative Result.} Combining all the above results, we get the perturbative partition function:
\equ{Z^{pert}_{(\wt{S^2_{ε_1}}×S^1)×Σ_{\fg_2}}=\frac{1}{|W_G|}∑_{\fm_1,\fm_2}∮d\tu\,\wt{H}^{\fg}\,e^{\frac{16π^3}{g^2}\Tr(\fm_1\fm_2) +2πik \Tr(\fm_1\fm_2\tu)} \Z^{1-loop,vec}_{(\wt{S^2_{ε_1}}×S^1)×Σ_{\fg_2}}\Z^{1-loop,hyp}_{(\wt{S^2_{ε_1}}×S^1)×Σ_{\fg_2}}\,,
}
where $\wt{H}$ is the Hessian appearing in the handle gluing operator $\wt{\H}$. Comparing with \eqref{jkformula}, we can identify the following operators (we use $\wt{\hphantom{x}}$ here to distinguish similar operators of the previous section) 
\begingroup
\allowdisplaybreaks
\eqs{\wt{Π}^{pert}_a &=e^{\frac{16π^3}{g^2}\fm_1^a +2πik (\fm_1\tu)^a}∏_α\left[\frac{Z_B\Big(x^{\alpha}\fq_{1}^{1+\frac{α(\fm_1)}{2}};\fq_{1}\Big)}{Z_B\Big(x^{\alpha}\fq_{1}^{-\frac{α(\fm_1)}{2}};\fq_1\Big)}\right]^{α^a}∏_ρ\left[\frac{Z_B\Big(-x^{ρ}y\fq_{1}^{\frac{1+ρ(\fm_1)}{2}};\fq_{1}\Big)}{Z_B\Big(-x^{ρ}y\fq_{1}^{\frac{1-ρ(\fm_1)}{2}};\fq_1\Big)}\right]^{-ρ^a} \label{Ptaop}\\
\wt{Π}^{pert}_i &= ∏_ρ\left[Z_B\Big(-x^{ρ}y\fq_{1}^{\frac{1+ρ(\fm_1)}{2}};\fq_{1}\Big)·Z_B\Big(-x^{ρ}y\fq_{1}^{\frac{1-ρ(\fm_1)}{2}};\fq_1\Big)^{-1}\right]^{-1} \label{Ptiop}\\
\wt{\H}^{pert} &=∏_α\left[Z_B\Big(x^{\alpha}\fq_{1}^{1+\frac{α(\fm_1)}{2}};\fq_{1}\Big)·Z_B\Big(x^{\alpha}\fq_{1}^{-\frac{α(\fm_1)}{2}};\fq_1\Big)^{-1}\right]^{-1}\det_{ab}\frac{∂\log(Π^{pert}_a)}{2πi ∂\tu_b} \nn
&=∏_{α>0}\left[(-1)^{-α(\fm_1)}x^α\Big(1-\fq_1^{-\frac{α(\fm_1)}{2}}x^{-α}\Big)\Big(1-\fq_1^{\frac{α(\fm_1)}{2}}x^{-α}\Big)\right]^{-1}\det_{ab}\frac{∂\log(Π^{pert}_a)}{2πi ∂\tu_b}\,·
\label{Htop}}
\endgroup
The flavour magnetic fluxes $\fn_1$ can again be introduced by shifting $y→y\fq_1^{\frac{\fn_1}{2}}$ whenever needed. Of course, the perturbative Bethe vacua follows from
\equ{\wt{\S}^{pert}_{BE}=\{\hat{u}\;|\;\wt{Π}^{pert}_a(\hat{u})≡\text{exp}\(2πi\tfrac{∂\W^{pert}}{∂\tu_a}(\hat{u})\)=1,\quad a=1,⋯,r_G\}/W_G,
\label{BetheEq3dTTI}}
from where we find the perturbative effective twisted potential $\W^{pert}_{(\wt{S^2_{ε_1}}×S^1)×Σ_{\fg_2}}$. The classical contribution simply reads
\equ{\W^{cl}_{(\wt{S^2_{ε_1}}×S^1)×Σ_{\fg_2}}=\frac{8π^2\fm_1·\tu}{i g^2} +\frac{k}{2}\Tr(\fm_1\tu^2)\,.
}
The 1-loop contributions to $\W$ involve integrating the function $\log Z_B$, so we can again use the $\check{T}$-functions defined in \eqref{defTcheck} to write the following contributions:
\begingroup
\allowdisplaybreaks
\eqs{(2πi)\W^{1-loop,vec}_{(\wt{S^2_{ε_1}}×S^1)×Σ_{\fg_2}} &=\frac{1}{2}∑_{α>0}\left[\begin{matrix} \check{T}\(α(\tu)+1+ε_1,\tfrac{ε_1α(\fm_1)}{2}|ε_1\) +\check{T}\(α(\tu),\tfrac{ε_1α(\fm_1)}{2}|ε_1\) \\ -\check{T}\(α(\tu)+1,\tfrac{-ε_1α(\fm_1)}{2}|ε_1\) -\check{T}\(α(\tu)+ε_1,\tfrac{-ε_1α(\fm_1)}{2}|ε_1\) \end{matrix}\right], \\
(2πi)\W^{1-loop,hyp}_{(\wt{S^2_{ε_1}}×S^1)×Σ_{\fg_2}} &=-\frac{1}{2}∑_{ρ}\left[\begin{matrix} \check{T}\(ρ(\tu)+\tnu+\tfrac{1+ε_1}{2},\tfrac{ε_1ρ(\fm_1)}{2}|ε_1\) \\ -\check{T}\(ρ(\tu)+\tnu+\tfrac{1+ε_1}{2},\tfrac{-ε_1ρ(\fm_1)}{2}|ε_1\) \end{matrix}\right].
}
\endgroup
Note that we have again simply assumed $ρ>0$ to write the above form of the hyper contributions. For explicit representations, one has to make sure that this is true by switching between the analogues of the two terms present in each line of the vector contribution.

After we find the Bethe vacua solutions from extremizing the twisted superpotential at large $N$, we can use this solution to write the final result for the partition function at large $N$ as follows:
\equ{Z_{(\wt{S^2_{ε_1}}×S^1)×\Sigma_{\fg_2}}(\tnu;\fn_1,\fn_2) ≈\sum_{\hat{u} \in \wt{\S}_{\mathit{BE}}} \wt{\Pi}_i(\hat{u},\nu)^{\fn_{2i}(1-\fg_2)} \wt{\H}(\hat{u},\nu)^{\fg_2-1}\,,
\label{ZtgenPHform}}
with the relevant operators explicitly given in \eqref{Ptiop} and \eqref{Htop}. Again, due to subleading nature of the instanton contributions at large $N$, we have dropped the specifier `\emph{pert}' above.

We can try to relate $\log Z$ to $\W$ here too as done in the previous section but we will refrain from doing that and jump directly to the evaluation of the $Π$'s and $\H$'s explicitly at large $N$.

\subsection[\texorpdfstring{Large $N$ Limit (mIIA Duals)}{Large N Limit (mIIA Duals)}]{Large $\bm{N}$ Limit (mIIA Duals)}
We choose here the same Ansatz as before, i.e., both $\tu$ and $\fm_1$ have the same large $N$ behaviour. Explicitly,
\equ{\tu^i =-iN^{α}x\,,\quad \fm_1^i =-iN^{α} η x\,,
}
where $η$ needs to be determined via its EoM (and it will not vanish in this case).

\paragraph{Seiberg Theory.} Let us work out this special example before giving the general result. First, the twisted superpotential
\begingroup
\allowdisplaybreaks
\eqs{(2πi)\W^{Seiberg}_{(\wt{S^2_{ε_1}}×S^1)×Σ_{\fg_2}} &=-\frac{π}{4}N^{\frac{5}{2}}\left[\frac{(8-N_f)\big(4+3ε_1η(ε_1η+2)\big)}{3ε_1}∫ρ(x)x^3dx \right. \nn
&\qquad\qquad \left. -(ε_1η+2)\frac{(1+ε_1)^2-(2\tnu +ε_1\fn_1)^2}{4ε_1}∫dxdyρ(x)ρ(y)|x±y|\right] \nn
&\quad +\frac{π}{4}N^{\frac{5}{2}}\left[\frac{(8-N_f)\big(4+3ε_1η(ε_1η -2))\big)}{3ε_1}∫ρ(x)x^3dx \right. \nn
&\qquad\qquad \left. +(ε_1η -2)\frac{(1-ε_1)^2-(2\tnu -ε_1\fn_1)^2}{4ε_1}∫dxdyρ(x)ρ(y)|x±y|\right] \nn
&=-π N^{\frac{5}{2}}\left[(8-N_f)η∫ρ(x)x^3dx \right. \nn
&\qquad \left.-\left\{\frac{1}{2}\(1-2\fn_1\tnu\) +\frac{1}{8}\(1+ε_1^2 -ε_1^2\fn_1^2 -4\tnu^2\)η \right\} ∫dxdyρ(x)ρ(y)|x±y|\right].
\label{WtttiIntg}}
\endgroup
The above expression now needs to be extremized with respect to both $η$ and $ρ(x)$. Let us compare \eqref{WtttiIntg} with \eqref{Fdef} to get $\tilde{F}=-(2πi)\W\frac{1}{π}N^{-\frac{5}{2}}$ with $n=3$, $A(η)=\frac{1}{2}\(1-2\fn_1\tnu\) +\frac{1}{8}\(1+ε_1^2 -ε_1^2\fn_1^2 -4\tnu^2\)η$, $B(η)=(8-N_f)η$. Then, using \eqref{soln} we get
\eqsc{\(\tfrac{1}{2}\(1-2\fn_1\tnu\) +\tfrac{1}{8}\(1+ε_1^2 -ε_1^2\fn_1^2 -4\tnu^2\)η\)(8-N_f) -3(8-N_f)η\tfrac{1}{8}\(1+ε_1^2 -ε_1^2\fn_1^2 -4\tnu^2\)=0 \nn
⇒η=\frac{2(1-2\fn_1\tnu)}{1+ε_1^2 -ε_1^2\fn_1^2 -4\tnu^2}\,·
}
This sets $A=\frac{3}{4}(1-2\fn_1\tnu)$ and using \eqref{solmu} in \eqref{solF}, leads to
\eqs{(2πi)\W^{Seiberg}_{(\wt{S^2_{ε_1}}×S^1)×Σ_{\fg_2}} &=-\(-πN^{\frac{5}{2}}\)\frac{2}{5}\frac{4^2\(\sqrt{\frac{3}{4}(1-2\fn_1\tnu)}\)^3}{\big(\sqrt{3}\big)^3\sqrt{(8-N_f)η}} \nn[1mm]
&=\frac{2\sqrt{2}π N^{\frac{5}{2}}(1-2\fn_1\tnu)\sqrt{1+ε_1^2 -ε_1^2\fn_1^2 -4\tnu^2}}{5\sqrt{8-N_f}} \nn
&=-\frac{2}{9}(1-2\fn_1\tnu)\sqrt{1+ε_1^2 -ε_1^2\fn_1^2 -4\tnu^2}\,F^{Seiberg}_{S^5}\,.
\label{W3dTTIsbg}}
This form is similar to \eqref{logZSbgexp} and can be related to $\log Z_{S^3×Σ_{\fg_2}}$ as follows:
\equ{2πi(\fg_2-1)\W^{Seiberg}_{(\wt{S^2_{ε_1}}×S^1)×Σ_{\fg_2}}(\tnu;\fn_1) =-\frac{1}{4}\log Z_{S^3×Σ_{\fg_2}}\Big(2\tnu±ε_1\sqrt{\fn_1^2-1};\fn_1\Big)\,.
\label{relWTTIZS3}}
The unrefined version of this relation has also appeared in \cite{Hosseini:2018uzp}. The above relation suggests we can write a `twisted superpotential' for the twisted superpotential; calling the former a ``prepotential'', we can indeed do that via relation 3 in Table \ref{tab:relPFs}:
\equ{\F^{Seiberg}_{(\wt{S^2_{ε_1}}×S^1)×Σ_{\fg_2}}(\tnu;\fn_1) =-\frac{1}{27}F_{S^5}\bigg(3\tnu±\frac{3}{2}ε_1\sqrt{\fn_1^2-1}\bigg) =-\frac{1}{27}\big(1+ε_1^2 -ε_1^2\fn_1^2 -4\tnu^2\big)^{\frac{3}{2}}F_{S^5}\,.
\label{prepotF}}
The twisted superpotential $\W$ can then be written in terms of the prepotential $\F$ (similar to \eqref{logZ3dSCIrelWgen}) as follows:
\equ{(2πi)\W^{Seiberg}_{(\wt{S^2_{ε_1}}×S^1)×Σ_{\fg_2}} =\left[6\F +\(\fn_{1} -2\tnu\)\frac{∂\F}{∂\tnu} -2ε_1\frac{∂\F}{∂ε_1}\right].
}

Finally, moving on to the partition function that follows from \eqref{ZtgenPHform}, we have
\begingroup
\allowdisplaybreaks
\eqs{\log Z_{(\wt{S^2_{ε_1}}×S^1)×Σ_{\fg_2}}^{Seiberg} &=-πN^{\frac{5}{2}}(\fg_2-1)\left[\(\tfrac{1+ε_1}{2} -\fn_2\(\tnu +\tfrac{ε_1\fn_1}{2}\)\)\tfrac{(ε_1η+2)}{2ε_1}\right. \nn
&\quad \left.+\(\tfrac{1-ε_1}{2} -\fn_2\(\tnu -\tfrac{ε_1\fn_1}{2}\)\)\tfrac{(ε_1η-2)}{2ε_1}\right]∫dxdyρ(x)ρ(y)|x±y| \nn
&=-πN^{\frac{5}{2}}(\fg_2-1)\left[(1-\fn_1\fn_2)+\frac{η}{2}(1 -2\fn_2\tnu)\right]∫dxdyρ(x)ρ(y)|x±y| \nn
&=\frac{4}{9}(\fg_2-1)\frac{(1-\fn_1\fn_2)(1+ε_1^2 -ε_1^2\fn_1^2 -4\tnu^2) +(1-2\fn_1\tnu)(1-2\fn_2\tnu)}{\sqrt{1+ε_1^2 -ε_1^2\fn_1^2 -4\tnu^2}}\,F_{S^5}^{Seiberg}\,.
\label{TTIsbg}}
\endgroup
This matches the result of \cite{Hosseini:2018uzp} in the $ε_1→0$ limit, as the terms proportional to $ε_1^2$ then vanish. The coefficient of the integral expression is split in such a way that again leads to factorization reminiscent of 3d Cardy formula \cite{Choi:2019dfu}, but only in the strict Cardy limit, as in the previous section.

\paragraph{Extremization.} We can also extremize \eqref{TTIsbg} with respect to $\tnu$ which helps in performing holographic checks. For generic flavour fluxes $\fn_1,\fn_2$, the extremization of \eqref{TTIsbg} leads to a cubic equation and the solutions are not too illuminating. Restricting ourselves to $\fn_1=\frac{1}{2\fn_2}$, leads to
\begingroup
\allowdisplaybreaks
\eqsc{\tnu^* =\frac{(1+2\fn_2^2)(4\fn_2^2 -ε_1^2(1-4\fn_2^2))}{2\fn_2(12\fn_2^2 -ε_1^2(1-4\fn_2^2))} \\
⇒\log Z^{Seiberg}_{(\wt{S^2_{ε_1}}×S^1)×Σ_{\fg_2}} =\frac{(\fg_2-1)}{9\fn_2}\sqrt{\frac{(1-4\fn_2^2)(4\fn_2^2-ε_1^2)\big(4(\fn_2^2-1)-ε_1^2(1-4\fn_2^2)\big)}{4\fn_2^2 -ε_1^2(1-4\fn_2^2)}}F^{Seiberg}_{S^5} \nn
\xrightarrow{ε_1→0}\frac{2}{9\fn_2}(\fg_2-1)\sqrt{(1-4\fn_2^2)(\fn_2^2-1)}F^{Seiberg}_{S^5}\,.
}
\endgroup
Note that requiring a real, nonzero, unrefined $\log Z$ above puts a constraint on the flavour fluxes: $\half<|\fn_{1,2}|<1$. It would be interesting to compare this with explicit holographic computation, along the lines of \cite{Bah:2018lyv}.

\paragraph{Generic Quivers.} As in the previous section, we state here the general results as the matrix model analysis is a straightforward extension of the Seiberg theory discussed above. The twisted superpotential and the partition function then read in terms of ``prepotential'' $\F$ as follows:
\begingroup
\allowdisplaybreaks
\eqsc{\F_{(\wt{S^2_{ε_1}}×S^1)×Σ_{\fg_2}}(\tnu_I;\fn_{1I}) =-\frac{1}{27}F_{S^5}\bigg(3\tnu_I±\frac{3}{2}ε_1\sqrt{\fn_{1I}^2-1}\bigg) \\
(2πi)\W_{(\wt{S^2_{ε_1}}×S^1)×Σ_{\fg_2}}(\tnu_I;\fn_{1I}) =\bigg[6\F +∑_I\(\fn_{1I} -2\tnu_I\)\frac{∂\F}{∂\tnu_I} -2ε_1\frac{∂\F}{∂ε_1}\bigg] \label{relWFTTI}\\
\log Z_{(\wt{S^2_{ε_1}}×S^1)×Σ_{\fg_2}}(\tnu_I;\fn_{1I},\fn_{2I}) =-2πi(\fg_2-1)\bigg[4\W +∑_I\(\fn_{2I} -2\tnu_I\)\frac{∂\W}{∂\tnu_I} -2ε_1\frac{∂\W}{∂ε_1}\bigg].
\label{logZ3dTTIrelWgen}}
\endgroup
In the unrefined limit $(ε_1=0)$, the last term vanishes so the above relation reduces to the expected large $N$ relation between topologically twisted index $\log Z_{\wt{S^2}×S^1}$ and $S^3$ free energy of 3d $\N=2$ theories \cite{Hosseini:2016tor,Jain:2019lqb}. To see this, we can rewrite \eqref{relWFTTI} following \eqref{relWTTIZS3} as
\equ{2πi(\fg_2-1)\W_{(\wt{S^2_{ε_1}}×S^1)×Σ_{\fg_2}}(\tnu_I;\fn_{1I}) =\frac{1}{4}F_{S^3×Σ_{\fg_2}}\Big(2\tnu_I±ε_1\sqrt{\fn_{1I}^2-1};\fn_{1I}\Big)\,.
\label{relWFS3}}

We can further compactify the above expressions by choosing a quartet of constrained flavour fugacities and fluxes for each parameter in the trio $\{\tnu_I,\fn_{1I},\fn_{2I}\}$ as follows:
\begingroup
\allowdisplaybreaks
\eqsc{\tnu^{+±}_I =(1+ε_1)±(2\tnu_I +ε_1\fn_{1I})\,,\quad \tnu^{-±}_I =(1-ε_1)±(2\tnu_I -ε_1\fn_{1I})\,, \\
\fn_{iI}^{+±}=2(1±\fn_{iI})\,,\quad \fn_{iI}^{-±}=2(1±\fn_{iI})\quad\text{ for }i=1,2\,. \\
\F_{(\wt{S^2_{ε_1}}×S^1)×Σ_{\fg_2}} =-\frac{1}{27}\(\frac{{\textstyle ∑_I}c_{ρ}^I\big[\tnu_I^{++}\tnu_I^{+-} +\tnu_I^{-+}\tnu_I^{--}\big]}{2∑_Ic_{ρ}^I}\)^{\frac{3}{2}}F_{S^5} \\
(2πi)\W_{(\wt{S^2_{ε_1}}×S^1)×Σ_{\fg_2}} =∑_{\{±\},I}\fn_{1I}^{±±}\frac{∂\F}{∂\tnu_I^{±±}} \\
\log Z_{(\wt{S^2_{ε_1}}×S^1)×Σ_{\fg_2}} =-(2πi)(\fg_2-1)∑_{\{±\},I}\fn_{2I}^{±±}\frac{∂\W}{∂\tnu_I^{±±}} =-(\fg_2-1)∑_{\{±\},I,J}\fn_{1I}^{±±}\fn_{2J}^{±±}\frac{∂^2\F}{∂\tnu_I^{±±}∂\tnu_J^{±±}}\,·
\label{logZ3dTTIrelWgenComp}}
\endgroup
These relations for refined index of generic quiver theories generalize the unrefined index of Seiberg theory appearing in \cite{Hosseini:2018uzp}.

\paragraph{Universal.} Finally, let us set all flavour chemical potentials and fluxes to zero. The following relations then hold for generic quivers:
\eqsg{(2πi)\W^{univ}_{(\wt{S^2_{ε_1}}×S^1)×Σ_{\fg_2}}=-\frac{2}{9}\sqrt{1+ε_1^2}\,F_{S^5} \\
\log Z^{univ}_{(\wt{S^2_{ε_1}}×S^1)×Σ_{\fg_2}}=\frac{4}{9}(\fg_2-1)\left[\sqrt{1+ε_1^2} +\frac{1}{\sqrt{1+ε_1^2}}\right]F_{S^5}\,.
}
The latter relation reduces correctly in the unrefined limit to
\equ{\log Z^{univ}_{(\wt{S^2}_{ε_1=0}×S^1)×Σ_{\fg_2}} =\frac{8}{9}(\fg_2-1)F_{S^5} =-F_{S^3×Σ_{\fg_2}}\,.
}
The final equality can be obtained by combining \eqref{logZ3dTTIrelWgen} and \eqref{relWFS3} (or refer to \cite{Crichigno:2018adf,Hosseini:2018uzp}) and is as expected for 3d theories. It would be interesting to verify the general refined results derived above with explicit results from the gravity side but to the best of our knowledge, such holographic computations do not exist yet.

\section{Outlook}\label{sec:Disc}
In this note, we computed partition functions of 5d $\N=1$ gauge theories on $(S^2_{ε_1}×S^1)×Σ_{\fg_2}$ without and with a topological twist on $S^2$. We also turned on the refinement $ε_1$ for $S^2$ corresponding to angular momentum. The large $N$ results we have obtained for these indices extend the applicability of 3d Cardy formulae of \cite{Choi:2019dfu} to the novel class of 3d theories obtained from compactification of 5d $\N=1$ SCFTs. This class of theories have partition functions scaling as $N^{\frac{5}{2}}$ in the large $N$ limit compared to the partition functions of the usual 3d theories with M-theory duals that scale as $N^{\frac{3}{2}}$. Also, it seems the Cardy factorization associated with the latter class of 3d theories with finite $ε_1$ does not hold for the class of 3d theories obtained from 5d; the factorization holds only in the strict Cardy limit $ε_1≪1$. This is in contrast to the expressions in \cite[Sec. 7]{Hosseini:2020wag}, which predict the same 3d factorization to hold in 5d too\footnote{We also do not seem to get the correct universal limit from the expressions given there.}. These partition functions are useful observables allowing one to explore physics of field theories in various dimensions, see for example \cite{Crichigno:2018adf,Hosseini:2018uzp,Razamat:2019sea,Sacchi:2021afk}. We expect that these refined partition functions will positively increase the tools available to further this exploration. We discuss a few applications along these lines below.

One of the most accessible extension of the results discussed here is the application to $\N=2$ super Yang-Mills theory along the lines of \cite{Crichigno:2018adf}. This theory is expected to have a UV completion as the circle compactification of a 6d $\N=(2,0)$ SCFT. Thus, the $\M_3×Σ_{\fg}$ partition function one computes for $\N=2$ SYM can be expected to compute the $\M_3×Σ_{\fg}×S^1$ partition function of the 6d theory. Changing the perspective slightly, it would be the same as computing $\M_3×S^1$ partition function of a 4d theory obtained by compactifying the 6d theory on $Σ_{\fg}$. That is, the 5d computation can be interpreted as giving the partition function of these 4d theories. Of course, this would be true after including all the instanton corrections in 5d. Thus, one should be able to compute generalized 4d indices like $(S^2_ω×S^1)×S^1$ with the setup discussed here. Practically, the instanton contributions may not be fully tractable in general, but in certain limits like the Schur limit, one can still compute exact and complete results in 4d, for which few other direct methods exist. This approach should work for other choices of $\M_3$ like Lens spaces or $Y^{p,q}$'s or generic Seifert manifolds too, which will allow access to more generic 4d indices. We plan to address some of these issues in future work.

For a given $\M_3$, the partition function of the 3d theory obtained by reduction on $Σ_{\fg}$ is computed by an appropriate 2d TQFT, in what may be called a ``3d-2d correspondence''. The large $N$ results suggest that a large class of novel 3d $\N=2$ SCFTs exist, arising as the IR fixed point of 5d quiver theories (including the orbifolded Seiberg theories), compactified on a Riemann surface. These are labeled by the discrete flavor fluxes $\fn$ and their indices (free energy) scales as $N^{\frac{5}{2}}$. It would be worth investigating whether some of these theories can be understood purely in terms of simpler building blocks of three-dimensional theories. Some steps have been taken towards this goal in \cite{Razamat:2019sea,Sacchi:2021afk}.

\

The unrefined 3d SCI and TTI have been well-studied for various theories but the same can not be said about the refined indices. So the large $N$ analysis of refined indices in 3d directly would be a welcome development, which can help verify and/or contrast the 3d consequences of the 5d results derived here.

The $AdS_6$ black hole solutions dual to the field theories discussed here with angular momentum refinement and generic flavour fluxes have not been explicitly constructed yet. The Bekenstein-Hawking entropy of such charged, rotating black holes would be determined by the large $N$ limit of the corresponding indices computed in this note. The field theory computations might also be suitably modified to include subleading corrections, which could allow for nontrivial tests of future holographic computations.

\section*{\centering Acknowledgements}
DJ thanks P.M. Crichigno for collaboration at the initial stages of this project along with many insightful and helpful discussions. DJ also thanks A. Manna for making accessible the identities given in Appendix \ref{app:Id} and more.

\appendix
\section{Special Functions}\label{app:Id}
\paragraph{$\bm{q}$-Pochhammer symbol (finite case).} It is defined by
\equ{(x;q)_n=∏_{i=0}^{n-1}(1-xq^i)\,.
}
Some relevant identities to simplify expressions in the main text are as follows:
\eqs{(q^k x;q)_n &=\frac{(x;q)_n(q^n x;q)_k}{(x;q)_k}\,, \\
(q^{-k} x;q)_n &=\frac{(x;q)_n(qx^{-1};q)_k}{(q^{1-n}x^{-1};q)_k}q^{-kn}\,, \\
(x;q)_n &=(q^{1-n}x^{-1};q)_n(-x)^nq^{\frac{n(n-1)}{2}}\,.
}
We always assume $k,n>0$.

\paragraph{$\bm{q}$-Pochhammer symbol (infinite case).} It is defined by
\eqs{(x;q)_{∞} &= \prod_{k\geq 0}(1-x q^{k})\,\qquad |q|<1 \\
&=∑_{n=0}^{∞}\frac{(-1)^nq^{\frac{1}{2}n(n-1)}}{(q;q)_n}x^n\,.
}
The summation form is valid for both $|q|<1$ and $|q|>1$. Some relevant identities are
\begingroup
\allowdisplaybreaks
\eqs{(x;q)_∞ &=\frac{1}{(q^{-1}x;q^{-1})_∞}\,, \\
(q^k x;q)_∞ &=\frac{(x;q)_∞}{(x;q)_k}\,, \\
(q^{-k}x;q)_∞ &=\frac{(-x)^k x q^{-\frac{k(k+1)}{2}}}{x-1}(x^{-1};q)_{k+1}(x;q)_∞\,.
}
\endgroup
We have suppressed the $∞$ subscript in the main text.

\paragraph{$\bm{q}$-theta function.} It can be defined in terms of the function given above as follows:
\equ{Θ(x;q)=(x;q)_∞(qx^{-1};q)_∞\,.
}
We will use the exponentiated parameters $x=e^{2πiz}$, $q=e^{2πiω}$ interchangeably in the arguments of functions in what follows. We note the following identity that relates $q$-theta function to Barnes' double gamma function $Γ_2(z|\vec{ω})$ \cite{FRIEDMAN2004362}
\equ{Θ(z;ω)^{-1} =e^{πiT(z|ω)}Γ_2(z|1,ω)Γ_2(1-z|1,-ω)Γ_2(1+ω-z|1,ω)Γ_2(z-ω|1,-ω)\,,
\label{qThExp}}
where $T$ is the quadratic polynomial
\equ{T(z|ω) =\frac{z^2}{ω} -\frac{ω-1}{ω}z +\frac{ω^2+3ω-1}{6ω}\,·
}
From \cite{RUIJSENAARS2000107}, we can read off the asymptotics of $Γ_2$ function as $|z|→∞$
\eqs{\log Γ_2(z|ω_1,ω_2) &≈ -\frac{1}{2!}B_{2,2}(z|ω_1,ω_2)\log z +∑_{n=0}^{1}\frac{B_{2,n}(0|ω_1,ω_2)z^{2-n}}{n!(2-n)!}∑_{l=1}^{2-n}\frac{1}{l} +⋯ \nn
&≈-\frac{1}{2}\(\frac{z^2}{ω_1ω_2} -\frac{ω_1+ω_2}{ω_1ω_2}z +\frac{ω_1^2+3ω_1ω_2+ω_2^2}{6ω_1ω_2}\)\log z +\frac{3}{4ω_1ω_2}z^2 -\frac{ω_1+ω_2}{2ω_1ω_2}z\,.
\label{logG2exp}}
Plugging \eqref{logG2exp} in \eqref{qThExp} and choosing the branch $\log(-u)=-iπ +\log(u)$, we get the large $z$ behaviour of $q$-theta function
\equ{\log Θ(z;ω)^{-1} ≈ πiT(z|ω)-\half\,.
\label{logThetaId}}
This relation is used extensively in the main text.

\references{5dRefs}

\end{document}